\documentclass[12pt, a4paper]{article}

\usepackage{amsthm, amsfonts, amssymb, braket, framed, graphicx, fancyhdr, eurosym, url, multirow}
\usepackage{amsmath}
\usepackage{subcaption}
\usepackage[top=30mm, bottom=30mm, right=25mm, left=25mm, headheight=15.72pt]{geometry}
\usepackage[utf8]{inputenc}
\usepackage[british,UKenglish,USenglish,american]{babel}
\usepackage{mathabx}
\usepackage{mathtools}

\usepackage{cite} 
\usepackage{hyperref}
\usepackage{upgreek}
\usepackage{comment}
\usepackage{mathrsfs}
\usepackage[dvipsnames]{xcolor}
\usepackage{tikz}

\usepackage[overload]{empheq}
\usepackage{comment}
\usepackage{physics}
\usepackage{url}
\newtheorem{theorem}{Theorem}

\newtheorem{exercise}{Exercise}
\usepackage{pgfplots,amsmath}
\pgfplotsset{compat=1.17} 
\usepackage{textcomp} 

\numberwithin{equation}{section}

\title{An introduction to Goldstone boson physics and to the coset construction}
\date{}
\author{Daniel Naegels$^{a,}$\footnote{daniel.naegels@ulb.be}}

\begin{document}

\maketitle

\begin{center}\it{
$^{a}$Physique Th\'eorique et Math\'ematique and International Solvay Institutes, \\ Universit\'e Libre de Bruxelles, C.P. 231, B-1050 Brussels, Belgium\\}
\end{center}
\vspace{25pt}

\begin{abstract}
\noindent
These lecture notes are based on a six-hour series of lectures given at the XVII Modave summer school in
mathematical physics, aimed at Ph.D. students in high-energy theoretical physics. \\
\\
\noindent
The manuscript starts by briefly stating Goldstone's theorem and emphasises the motivations behind Goldstone physics; the main asset being the universality of spontaneous symmetry breaking (SSB) which is the fundamental hypothesis of Goldstone's theorem. Once the different notions of SSB will be clarified/reviewed, Goldstone's theorem will be stated and proved. A prediction of this theorem is the existence of gapless particles, called Nambu-Goldstone modes (NG modes). From the discussion on Goldstone's results, some aspects of the NG modes will emerge. Besides to be gapless, they are systematically weakly coupled at low energy. Therefore, an effective field theory (EFT) building tool called ``coset construction'' will be presented to explicitly display these specific features of the NG modes. The coset construction suits our goal since it is mainly based on the symmetry realisation of the perturbed theory around the background inducing SSB. From the general obtained EFT, a counting rule for the NG modes will be derived. The limitations of this rule as well as the still ongoing generalisation will be discussed (e.g. spacetime symmetry breaking). The tools developed during this course will be illustrated with a concrete example in physics: ferromagnetism. The notes end with a brief state of the art of Goldstone physics. This provides some directions into which the interested reader could investigate to expand his knowledge on the subject.
\\
\\
\noindent
\emph{N.B.} : No prerequisites are required beside the standard courses of a Master in theoretical physics.

\end{abstract}
\newpage

\setcounter{tocdepth}{2}
\tableofcontents

\section{Introduction}
\label{Introduction}

\begin{figure}[!ht] 
 \begin{center}
  \includegraphics[scale=1.3]{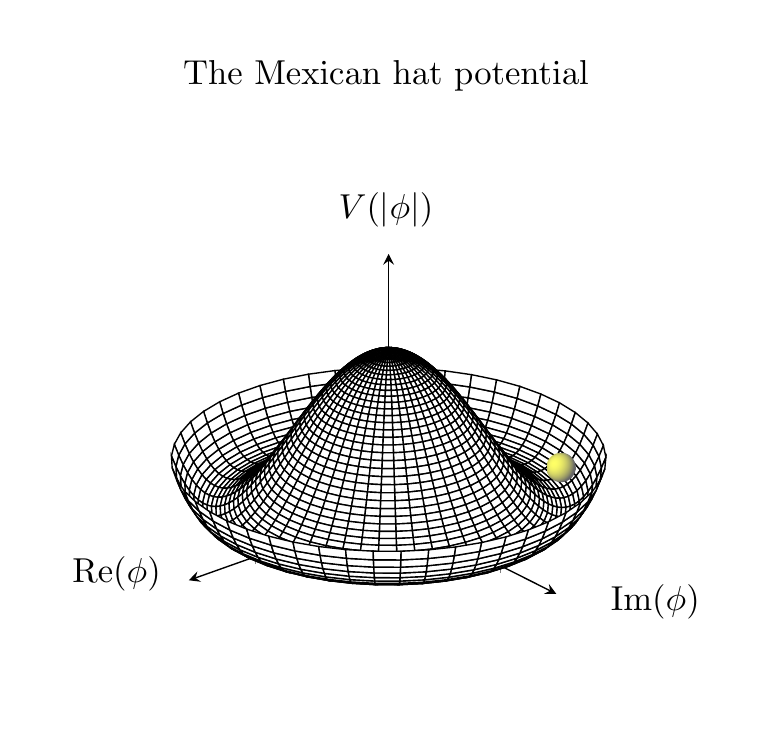}
 \end{center}
 \vspace*{-10mm}
 \caption{The dynamics of a physical system is mainly dictated by its potential. We observe that the represented potential in the above cartoon has a rotation symmetry since all the directions in the complex plane are equivalent. This means that the theory describing our system possesses this symmetry. However, the stable state into which we place the system (represented by the yellow sphere) selects one specific direction. Hence, all the directions are not anymore equivalent, the rotation symmetry is broken by the state of the system. We say that the symmetry is spontaneously broken.}
   \label{MexicanHat}
\end{figure}

These lecture notes are meant as a first introduction to Goldstone physics. This area of physics revolves around the Goldstone theorem which, briefly stated, ensures there is at least one massless mode, called the NG mode\footnote{NG mode stands for Nambu-Goldstone mode. Historically, Nambu is the one who conjectured the link between symmetry breaking and the mass constraint it implies \cite{Nambu:1960tm, Nambu:1961tp}, while it is Goldstone who clarified and proved this conjecture \cite{Goldstone:1961eq, Goldstone:1962es}.}, in the spectrum of the theory when a spontaneous symmetry breaking (SSB) occurs. SSB is when the theory possesses a symmetry (i.e. the dynamics is invariant under the action of the symmetry) but the vacuum around which we perform the perturbation analysis/the quantisation does not have such symmetry. A cartoon visualisation of this concept can be observed at Figure \ref{MexicanHat} through the Mexican hat potential. These different notions will be revised more formally later. However, we can already understand that Goldstone's theorem provides some information on the spectrum content at low energy (the IR region). Therefore, combining this information with tools to build Effective Field Theories (EFT) would provide us an almost complete description of the IR physics. That could be a definition of what Goldstone physics is : thorough analysis of the Goldstone theorem and of the related results supplemented by/completed with EFT tools.  

The notes are structured as follow. We will begin by motivating the subject. Afterwards, the formalism we will use will be settled. It should be seen as a way to set the conventions and the definitions rather than an axiomatic goal. Furthermore, it will permit to review the necessary background knowledge. Then, Goldstone's theorem will be stated and proved (without any claim of full mathematical rigour). From a discussion on the NG modes, some of their properties will emerge: they are massless, they are weakly coupled in the IR and they transform non-linearly under the spontaneously broken symmetries. These properties will formally be displayed through the construction of a generic EFT for NG modes. This, by using the coset construction formalism, which will itself be introduced. Furthermore, the EFT approach will allow us to acquire additional knowledge compared to the prediction of Goldstone's theorem, namely, we will obtain a classification and a counting rule for the NG modes. The range of validity of these results will as well be detailed. The abstract discussions and developments we made so far will be followed by a concrete example in condensed matter: ferromagnetism. Finally, a brief state of the art of Goldstone physics will be provided. 

This set of lectures given at the XVII Modave Summer School in Mathematical Physics remains globally an introduction to Goldstone physics. The interested reader can extend his knowledge in this field through the references provided along these notes. The four main references that were used to write these lecture notes are \cite{Beekman:2019pmi, Brauner:2010wm , Burgess:1998ku,Watanabe:2019xul}. Let us mention that the subject of spontaneous symmetry breaking is vast and many articles appeared in the past decades. The bibliography of this work is not meant to be exhaustive, it focuses on the references the author is the most familiar with. We apologise for any unintentional omissions of relevant papers or reviews. 
\\
\\
\noindent
\emph{N.B.} : Let us mention that Sections \ref{Introduction} to \ref{Counting rule through coset construction} included were explicitly presented during the 6-hour lecture while Sections \ref{Concrete example : ferromagnetism} and \ref{Some further directions} were briefly commented. We however provide additional developments to the later sections in order to make these notes self-contained.  

\subsection{Motivations}

To motivate the study of Goldstone physics we need first to recap the well-known assets of symmetries. This, in order to put into perspective the interesting aspects of spontaneous symmetry breaking. 

It is intuitive that to know a symmetry of an object (geometric figures, mathematical equations etc. left invariant after a given transformation) permits to ease the description and the manipulation of the considered object. In physics, this idea has been formalised through the Noether theorems which establish a connection between symmetries and conserved quantities. A noteworthy point is that these conserved quantities are exactly conserved no matter how complex the dynamics is. Hence, symmetries offer exact (i.e. non-pertubative) results. Furthermore, symmetries rely on the mathematical description of physics, it is thus not specific to a given physical scenario\footnote{An example is that we use energy conservation in any area of physics.}. When a concept applies to several physical phenomena, we say that this concept is universal. This is the case of symmetries. Finally, the symmetries permit to constrain the shape of a Lagrangian when we do model building. The importance of symmetries in physics can be heuristically shown by noticing that symmetries are one of the current paradigm of modern physics: special relativity has for cornerstone Lorentz symmetry, general relativity is based on diffeomorphism invariance and the standard model is constructed on the notion of gauge symmetries.      

Paradoxically, physics is richer when symmetries are spontaneously broken. At first, we could think that we lose universality but this is not the case since many physical phenomena take place around a pre-existing background which breaks spontaneously the fundamental symmetries. For example, the crystal structure in solid state physics is usually taken as granted and it breaks Poincaré group (spatial translations and rotations as well as the boosts). Another example could be the quark condensate, at low energy the quarks are not free and form bound states. This condensation breaks the chiral symmetry. An additional possible false idea we could have when SSB occurs is that we lose the exactness of the results related to symmetries. This can be denied thanks to Goldstone's theorem where the massless aspect of the NG modes is exact. Furthermore, we can mention that even in the case of an explicit symmetry breaking, if we are able to write down the source generating the breaking, the Ward-Takahashi identities can be generalised and still be exact (see for example \cite{Amoretti:2016bxs,Argurio:2015wgr}). This can be relevant for pseudo NG modes, cf. Subsection \ref{Pseudo Goldstone modes}. Finally, the spontaneously broken symmetries still constrain the shape of the Lagrangian. However, it is less easy to see because these symmetries are now ``hidden''. We will see that the formal explanation is that they are non-linearly realised instead of being linearly realised\footnote{In physics, in general, the usual symmetries ($U(1)$, $SO(3)$, $SU(N)\ldots$) are realised through matrices acting on fields which makes their action intrinsically linear.} which makes the invariance of the Lagrangian less obvious. Gathering all these observations, we can (with a bit of exaggeration) say that spontaneous symmetry breaking gives us the IR matter content of a given physical phenomenon through Goldstone's theorem and it constrains the shape of the associated EFT. In other words, it completely settles the effective field theory. Therefore, while symmetries are giving partial information on the dynamics through the conserved quantities, spontaneous symmetry breaking provide all the dynamics at low energy. It is in this sense that physics is richer in the case of SSB. Of course, that is hasty said, it should not be taken literally but more as a guideline which motivates the study of Goldstone physics. 

Since the last assertion is the main motivation of Goldstone physics, it could be interesting to have an additional viewpoint on it -- in order to double check the consistency of this remark. It can be done through the concept of RG flow. In the UV (high energy) each physical phenomenon is described by one theory. When we follow the RG flow toward the IR, the irrelevant operators become progressively suppressed. We thus remain with a handful numbers of theories -- the number of parameters is now limited -- which are constrained by the symmetries. Indeed, the RG flow modifies the theories consistently with the symmetries (if there are no anomalies). Hence, in the IR, one theory is describing several physical phenomena. The effective field theories are therefore universal and symmetrically constrained. The different physical phenomena are discriminated by the different interpretation we give to the parameters (the mass, the compression modulus etc.) and by the numerical values of these parameters\footnote{An EFT is the most general theory we can write which is invariant under a given symmetry group and that is written as an expansion in energy (usually a power series in derivatives). The experimental precision (or just the desired precision) tells us where to truncate this series. From the UV theory or from experiments we are thus able to determine the remaining parameters and our EFT becomes predictive at low energy.}. Furthermore, when we go to low energy, the system tends to condense (e.g. liquid-solid phase transition at low temperature, quark condensate at low energy, Bose-Einstein condensation at low temperature). The condensate will spontaneously break some symmetries. We can thus apply Goldstone's theorem to have information on the IR spectrum content. This leads to the rough idea that IR physics is universally described by Goldstone physics. As mentioned earlier, it should be understood more as an argument motivating the subject rather than a strong statement. 

Till now, we used abstract ideas to justify the universality of spontaneous symmetry breaking and Goldstone physics. We will close this section by stating some concrete examples where such concepts are found (it is not an exhaustive list).
\begin{itemize}
	\item High energy physics: light mesons physics \cite{Burgess:1998ku}, composite Higgs model \cite{Gripaios:2015qya}, Higgs mechanism \cite{Rubakov:2002fi}. 
	\item Statistical physics: phase transitions (ferromagnetism \cite{Burgess:1998ku}, $\ldots$), transport phenomena (superfluidity \cite{Schmitt:2014eka}, $\ldots$), condensed matter (crystal structure, $\ldots$) \cite{Ashcroft:1976}. 
	\item Astrophysics: stellar superfluids (e.g. in neutron stars) \cite{Schmitt:2014eka}
\end{itemize}
Of course, these domains are interconnected and it is one of the reasons why spontaneous symmetry breaking occurs in many areas of physics. For example, spontaneous symmetry breaking intervenes in the study of neutron stars because the latter have the right thermodynamic conditions to sustain a superfluid phase. Some of the superfluid phase transitions correspond to a Bose-Einstein condensation which can be described by the spontaneous symmetry breaking of a $U(1)$ symmetry  \cite{Schmitt:2014eka}. Thus, with the single example of superfluid, SSB intervenes in phase transition physics, in transport phenomena study and in astrophysics.

\subsection{Goals of the lecture}

As we have seen from the motivations, Goldstone physics is a vast subject which cannot be covered in detail in a 6-hour series lecture. We should thus settle the aim of the lecture (and so, of these lecture notes). The canonical goal is of course to get the gist of what Goldstone physics is. It is also interesting to acquire some technical knowledge that can be re-used in other domains. To do so, we decided to focus on the coset construction which is a building tool for effective fields theories. To understand the subtleties and the technical difficulties of a given subject, it is a good practice to do some concrete proofs of already established results. Therefore, we will prove (with some shortcuts) two main results of Goldstone physics: the Goldstone theorem and one of the existing counting rules for the NG modes (Goldstone's theorem predicts the existence of NG modes but does not say how many of them there are). Finally, the main purpose of this lecture is that a non-expert of the field can acquire enough background knowledge to not be confused if she/he comes across spontaneous symmetry breaking related topics during her/his own literature reading (and provide her/him enough references such that she/he knows where to look if a deeper knowledge is needed). Indeed, it is not unlikely that it happens due to the universal aspect of SSB. 

\section{Setting the formalism}
\label{SettingtheformalismSection}

Before starting to discuss and computing physical quantities related to spontaneous symmetry breaking we will do some recap on the necessary prerequisites as well as establishing the framework we are going to work with. This is the purpose of this section. Let us mention that we do claim a scientific approach (i.e. justify every step and keep an appropriate rigour level) however, we do not claim a full mathematical rigour and we do not pretend to have an axiomatic approach of physics. This section should be understood as a way to refresh memory on standard notions (no claim of originality) and to establish the vocabulary as well as the definitions. 

We are going to work with physical quantum field theories (QFT) defined on Minkowski spacetime of dimension $d\geq 2$. Using fields as degrees of freedom is consistent with the infra red limit because this limit is equivalent to probe for phenomena occurring at large distances and so, there is not much loss of generality by considering the continuous limit (e.g. in crystal structure) \cite{Burgess:1998ku}. Furthermore, we need at least one spatial direction in addition to the time direction to be able to define a notion of momentum and of energy (necessary for the concept of mass/gap). The term ``physical'' is deliberately vague\footnote{The idea of physical theories is intrinsically vague since this notion evolves with our understanding of nature. A naive example is that at a moment of History we thought that time was absolute but with special relativity, we learned that it was not correct. So, what a physical theory is evolves through science History. A more relevant example for us is that in the early seventies Coleman stated that for a relativistic field theory in two spacetime dimensions, no spontaneous symmetry breaking can occur \cite{Coleman:1973ci}. This theorem can be evaded if we consider strictly large N theories \cite{Coleman:1974jh,Gross:1974jv,Witten:1978qu}. These theories could be thought as purely exotic and non-physical. However, thanks to the holographic duality postulated in the nineties \cite{Maldacena:1997re, Witten:1998qj, Gubser:1998bc}, it appears that they could be linked to consistent physical gravitational theories.} but it should include at least:
\begin{itemize}
	\item a notion of locality: following the theorems or the results we will consider, there will be specific restrictions on locality. For example, while building effective field theories, we will only consider interactions between fields evaluated at the same spacetime position. On the contrary, Goldstone's theorem is robust with respect to locality. It is valid up to interactions with a finite range in space. 
	\item stability: we want our vacuum as well as the fluctuations around it to be stable, i.e. to remain finite through time.
	\item Consistency with a possible Poincaré-relativistic UV completion: Goldstone physics encompasses phenomenological descriptions of macroscopic systems. These phenomenological theories could be non-relativistic, however, we know that at the fundamental level, physics is relativistic. Therefore, any non-relativistic theory in the IR should, at higher energy, comes from the spontaneous symmetry breaking of Poincaré symmetry (as it is suggested in \cite{Nicolis:2015sra} for example). We will not check explicitly this last requirement concerning a possible relativistic completion, but we should keep in mind this physical constraint. Notice that in this course, when we mention relativity it is with respect to Poincaré symmetry (Poincaré/Lorentz relativity).  
\end{itemize}

It is important to notice that the considered QFTs are not necessary Lorentz invariant. Since this course is essentially addressed to high energy physics Ph.D. students (where relativistic QFTs are the norm), we can take some time to justify why we can be interested in non-relativistic theories. We have that macroscopic systems correspond usually to some fluctuations around a given condensate (e.g. solid state physics). The centre of mass of this condensate corresponds therefore to a preferential frame which is opposed to the paradigm of relativity: fundamentally, the laws of physics are the same in any (inertial) frame. Furthermore, the thermodynamic state of the system is given through a static computation with the probability weight $e^{-\beta H}$ where $H$ is the Hamiltonian. The interplay between the Hamiltonian and the Lorentz group is non-trivial, thus, thermodynamic states tend to break Lorentz invariance. Finally, the macroscopic systems are important for this lecture because Goldstone's theorem does apply to them as well. 

Even if we just mentioned the importance of statistical field theory, we are going to work with QFTs at zero temperature and at zero chemical potential. Of course, switching on temperature and a chemical potential is part of the research area of Goldstone physics but, from the (subjective) point of view of the author, it does not (pedagogically speaking) belong to an introduction. Let us mention that a comment about the finite density case will be made in Section \ref{Some further directions}. 

For the computation perspective, we are going to use the mostly minus metric $$(+,-,-,\ldots)\ ,$$ and the natural units $c=1=\hbar$, where $c$ and $\hbar$ are, respectively, the speed of light and the Planck constant.

\subsection{What do we mean by symmetry ?}

\subsubsection*{At classical level}

In field theory, a symmetry is a transformation applied on the fields which leaves the equations of motion (EOM) unchanged. An equivalent formulation is that under such transformation, a solution of the EOM remains a solution. Mathematically, a transformation on the fields is defined as
\begin{equation}
    \begin{cases}
     \phantom{\phi^i(}x^{\mu} &\rightarrow \; x'^{\mu}=x'^{\mu}(x) \ , \\
     \phi^i(x) & \rightarrow \; \phi'^{i}(x')=F^{i}\left[x, \phi(x)\right] \ ,
    \end{cases}       
    \label{TrsfLaw}
\end{equation}
where $\phi$ is a generic field and the index $i$ refers to its possible multi-component nature, $F^{i}$ is a function, finally, the prime index represents the transformed object.

In this work we will do a small misnomer by defining ``a symmetry'' as a transformation applied on the fields which leaves the action of the field theory unchanged: 
	\begin{equation}
		S[\phi']=S[\phi] \ .
		\label{SymCl}
	\end{equation}
This small misappropriation of the term symmetry is consistent in the sense that a symmetry of the action implies a symmetry of the EOM. Furthermore, we do not lose much generality because most of the important symmetries (and the mostly used ones) in physics are the ones which could be seen at the level of the action. 

If we consider several transformations of the type \eqref{TrsfLaw}, we can combine them through the law of function composition and get another symmetry by ``chain reaction''. It is therefore possible to define an internal associative product. The identical transformation is trivially a symmetry. Finally, physical interesting transformations are predominantly invertible\footnote{A symmetry transformation could be somehow interpreted as a change of frame. Physically, nothing prevents us to go back to the original frame.} (e.g. rotations, phase-shifts, translations \ldots). Hence, the symmetries of a theory form a group. It is common to denote this set of transformations as a realisation of the usual groups: $\mathbb{Z}_2$, $U(1)$, $SO(3)$, $SU(N)$ \ldots. We will of course come back to it later, but Goldstone's theorem only applies when such groups are continuous. We will from now on focus on continuous groups. If we consider a continuous set of transformations \eqref{TrsfLaw} parametrised by $\alpha^a$, we can write an infinitesimal expression
\begin{equation}
    \begin{cases}
     \phantom{\phi^i(}x^{\mu} &\rightarrow \; x'^{\mu}=x^{\mu}+ \alpha^a \, \xi_a^\mu(x) \ , \\
     \phi^i(x) & \rightarrow \; \phi'^{i}(x)=\phi^i(x) + \alpha^a \, \delta_a\phi^i(x) \ .
    \end{cases}       
    \label{TrsfLawInfinit}
\end{equation}
These transformations correspond to the realisation of the continuous connected part to the identity ($\alpha=0$) of the symmetry group. Therefore, $\alpha^a$ parametrises the Lie algebra of the continuous group and we can define the representation of the generators $G_a$ by the infinitesimal action on the fields 
\begin{equation}
\alpha^a G_a \phi^i(x) \equiv \delta_\alpha \phi^i(x)\equiv \phi'^i(x)-\phi^i(x) \ .
\end{equation}
Let us mention that we will slowly start to stop to do the distinction between the generators and the realisation of the generators. The context should make it clear which case is considered. 

We recall that through Noether first theorem, it exists a one-to-one relation between the set of symmetry generators $G_a$ and the set of conserved currents $j_a^{\mu}(x)$ with $\partial_\mu j_a^{\mu} = 0$ on-shell. The conserved current associated to $G_a$ is constructed as follow \cite{Beekman:2019pmi, Peskin:1995ev}
\begin{equation}
j^\nu_a = \frac{\partial \mathcal{L}}{\partial \left( \partial_\nu \phi^i \right)}\delta_a\phi^i - K_a^\nu \ , 
\label{ConsCurrentWithBoundaryTerm}
\end{equation}
where $\mathcal{L}(\phi, \partial \phi)$ is the Lagrangian of the symmetric theory. This Lagrangian can transform up to a global derivative under the transformation 
\begin{equation}
\delta_a\phi^i \equiv G_a \phi^i  \ ,
\end{equation}
such that we express the global derivative through $K_a^\nu$:
\begin{equation}
\delta_a\mathcal{L}=\partial_\mu K^\mu_a \ .
\end{equation}
From $j_a^\mu$, a conserved quantity can be built
\begin{equation}
Q_a \equiv \int d^{d-1}x \,\, j_a^0(x) \ .
\end{equation}  

\subsubsection*{At quantum level}

Quantum mechanics is described by a complex Hilbert space $\mathcal{H}=\{\ket{\psi}\}$ and by a Hamiltonian. Conceptually, a symmetry transformation $\ket{\psi}\rightarrow \ket{\psi'}$ can be seen as a change of frame. Changing the frame should not alter the relative results of an experiment. Therefore, a necessary condition that a symmetry transformation should satisfy is   
\begin{equation}
\bra{\psi}\hat{A}\ket{\psi} = \bra{\psi'}\hat{A}'\ket{\psi'}\ ,
\label{NCSymQMHilbert}
\end{equation}
where $A$ is an observable and the prime represents its transformation. Wigner theorem states that for \eqref{NCSymQMHilbert} to be fulfilled, a symmetry transformation acting on $\mathcal{H}$ should be either unitary and linear or antiunitary and antilinear \cite{Weinberg:1995mt}. As it will become clear later, in Goldstone physics we are interested in the part of the symmetry group which is continuously connected to the identity. Said otherwise, we are interested in the symmetry transformations which can be parametrised by the Lie algebra. Identity is a unitary operator and the switch between unitary and antiunitary is discontinuous, thus, we have that our considered transformations are unitary. Hence, the symmetry transformations we are interested in, realising a given symmetry group, can be written as
\begin{equation}
\ket{\psi}\rightarrow e^{i \alpha^a \hat{Q}_a}\ket{\psi}  \ ,
\label{trsfLawMQ}
\end{equation}
where the realisation of the generators is Hermitian \begin{equation}
\hat{Q}_a^{\dagger}=\hat{Q}_a  \ .
\end{equation} 
 From \eqref{NCSymQMHilbert} and \eqref{trsfLawMQ}, we have
\begin{equation}
\hat{A} \rightarrow e^{i \alpha^a \hat{Q}_a} \hat{A} \, e^{-i \alpha^a \hat{Q}_a} \ .
\end{equation}

To define what a symmetry is, we will use a pragmatic approach. From the classical discussion we know that a symmetry is associated to a conserved quantity. Therefore, we are going to say that a continuous group of transformations $\ket{\psi}\rightarrow \ket{\psi'}$ realising on $\mathcal{H}$ one of the usual continuous groups ($SU(2)$, $O(N)$ etc.) is a symmetry group if the generators $\hat{Q}_a$ satisfy:
\begin{equation}
\frac{d \hat{Q}_a}{dt}\equiv \partial_t \hat{Q}_a + [\hat{Q}_a,H] = 0 \ .
\label{CSymQM}
\end{equation}

A final comment is that, from the canonical quantisation, the operatorisation of the conserved charges of a field theory corresponds to the realisation of the generators at quantum level 
\begin{equation}
[i\hat{Q}_a,\hat{\phi}^i]=\delta_a \hat{\phi}^i \equiv G_a \hat{\phi}^i \ ,
\label{quantisationConsCharg}
\end{equation}
if $[\delta_a \hat{\phi}^i,\hat{\phi}^j]=0$ \cite{Beekman:2019pmi}.

For the section \ref{Different classifications of the symmetries} and beyond, we will not denote with a circumflex accent anymore the quantum operators. 

\subsubsection*{\emph{A bit more of details}}

\emph{The phase space of a quantum theory corresponds to the projective space of $\mathcal{H}$ \cite{Weinberg:1995mt}. To say it more simply, a state of the system is a ray $\mathcal{R}$ of $\mathcal{H}$ (this because global phases are not observable). Let us consider a continuous set $\{T\}$ of transformations $T\, :\, \mathcal{R} \rightarrow \mathcal{R}'$ which has a group structure (through function composition) that realises one of the usual continuous groups ($SU(2)$, $O(N)$ etc.). From a transformation $T$ of our given set, we may define a transformation $U(T)$ acting on the Hilbert space, $U(T)\, :\,\ket{\psi}\rightarrow \ket{\psi'} $.The product law induced on $\left\lbrace U(T) \right\rbrace$ is defined up to a global phase (cf. the projective nature of the phase space vs. the vector space $\mathcal{H}$). Therefore, the representation of a given symmetry group on the phase space corresponds to a projective representation on the Hilbert space. It can be shown that a central charge might appear in the realisation of the Lie algebra on $\mathcal{H}$
\begin{equation}
[\hat{Q}_a, \hat{Q}_b] \sim f_{ab}^{\phantom{ab}c} \, \hat{Q}_c + c_{ab} \ ,
\end{equation}
where $f_{ab}^{\phantom{ab}c}$ are the structure constants \cite{Weinberg:1995mt}. Let us mention that to pursue with the manipulation of usual representations on $\mathcal{H}$, a possible trick is to consider a central extension of the symmetry group we want to realise on $\mathcal{H}$.  }
 
\subsection{Different classifications of the symmetries}
\label{Different classifications of the symmetries}

Goldstone physics depends heavily on symmetries, we can then naturally be convinced that some of the results we will state in the following sections rely on the nature of the considered symmetries. One way to characterise the symmetries is through the mathematical properties of the symmetry group. As we will see, the symmetry group being compact or not will play a major role. We then qualify the symmetries to be compact or non-compact. Other possible criteria on which to classify the symmetries could be the way the symmetries act on the fields and on spacetime. We list here the main important classifications: 

\begin{itemize} 
	\item Local symmetries are such that the parameters $\alpha^a$ in \eqref{TrsfLawInfinit} are functions of spacetime. Otherwise, we speak of global symmetries.
	\item Spacetime symmetries are symmetries which act non-trivially on spacetime. Consequently, a non-spacetime symmetry will have $x'^{\mu}=x^{\mu}$ in \eqref{TrsfLaw}. The typical examples of spacetime symmetries are translations, rotations, boosts etc. 
	\item Internal symmetries are the ones where the generators commute with the Poincaré algebra. It could be thought as being the non-spacetime symmetries. $U(1)$, $SO(2)$, $SU(N)$ \ldots are typical internal symmetries. 
	\item Uniform symmetries are the ones where $F^i$ in \eqref{TrsfLaw} does not depend explicitly on $x^{\mu}$. An equivalent definition is when the realisation of the generators does not depend on spacetime coordinates. For example, the generators of translations and rotations acting on a scalar field are respectively 
	\begin{align}
	   P_{i} = -\partial_{i} \ , \;
	   L_{ij} = x_i 	\partial_{j} - x_j \partial_{i} \ .
	\end{align}	
We notice that translations are uniform symmetries while rotations are not. Let us mention that translations are the only spacetime symmetries which are uniform. 
	\item Compact symmetries are the ones which realise compact groups.
\end{itemize}

\subsection{Spontaneous symmetry breaking}

Spontaneous symmetry breaking (SSB) is the phenomenon in which a stable state of the system transforms non-trivially under certain symmetries of the theory. These symmetries are then said to be spontaneously broken and the state is called the broken state \cite{Beekman:2019pmi}.

In classical field theory, the state of the system is characterised by one of the solutions of the EOM of the fields. We will call this particular solution the background, it can also be referred to as the vacuum. It is a stable solution if it remains finite along its evolution through spacetime and if, small perturbations around it remain small along their dynamical evolution. It is customary to look for such stable background among the solutions which minimise the energy, at least corresponding to a local minimum. This could be intuitively understood from point-like classical mechanics where the conservative forces act in the opposite direction to the gradient of the potential. Hence, being originally at a minimum of the potential, we have that the forces tend to bring back the system to its original state. Furthermore, by minimising the kinetic energy, we ensure that the system has not enough inertia to pass a potential hill. Otherwise, the system could go from one potential minimum to another one. 
	
At the quantum level, for a given symmetry, the vacuum state $\ket{0}$ of the system breaks spontaneously this symmetry if
\begin{equation}
e^{i \alpha Q}\ket{0}\neq\ket{0} \ \text{up to a global phase.}
\label{NaiveDefSSB}
\end{equation}
However, this naive definition of SSB might not be well settled because the non-trivial action of the broken generator $Q$ on our vacuum might lead to an ill defined state, i.e. a state with an infinite norm \cite{FabriE1966QFTa,Guralnik:1967zz}. Indeed, if $\ket{0}$ is homogeneous (i.e. an eigenstate of $P_\mu$) and $Q$ is uniform -- in addition to be Hermitian -- then 
\begin{align}
||Q\ket{0}||^2 & = \bra{0}Q^\dagger Q\ket{0} = \bra{0}QQ\ket{0} \ , \\
&= \int d^{d-1}x\bra{0}j^0(x)Q\ket{0} \ , \label{insideInt} \\
& = \bra{0}j^0(0)Q\ket{0}\int d^{d-1}x \label{outsideInt} \ ,
\end{align}
which, in infinite volume, could tend to infinity if $ Q\ket{0} \neq 0$. Notice that the symmetry being uniform, we have been able to express $j^0(x)$ as a translation in spacetime of $j^0(0)$,
\begin{equation}
j^0(x) = e^{i x^{\mu}P_\mu}j^0(0) e^{-i x^{\mu}P_\mu} \ .
\label{translationToOriginConsCur}
\end{equation}
By still using the uniform aspect of the symmetry, we have that $[Q,P_\mu]=0$ because $Q$ is either internal or a spacetime translation generator (we look to the case without central charges). Then, considering our vacuum as being homogeneous (we are not currently looking to the breaking of spacetime symmetries) and choosing it as the zero-energy (we are not considering gravity, only the relative energy among states is physical), we have $P_\mu\ket{0} = 0$. Combining these observations allowed us to go from \eqref{insideInt} to \eqref{outsideInt}.

A more formal definition of spontaneous symmetry breaking is then used to evade this possible inconsistency. We will say that a state $\ket{\psi}$ breaks the symmetry generated by $Q$ if there exists any field $\Phi$, called the interpolaing field, such that \cite{Watanabe:2019xul}:
	\begin{equation}
		\bra{\psi}[Q,\Phi(x)]\ket{\psi} \neq 0 \ .
		\label{DefSSBQMFormal}
	\end{equation}
If no such operator $\Phi$ exists, the state is symmetric. An argument to use a local field $\Phi$ to define SSB is that we are working in infinite volume or more generally in the thermodynamic limit (cf. the coming section about singular limits). It is thus more convenient to be able to probe locally if the SSB occured rather than to perform a global analysis on the full ket state \cite{Watanabe:2015eij}.

\begin{exercise}Show that \eqref{NaiveDefSSB} and \eqref{DefSSBQMFormal} are conceptually equivalent. \end{exercise}

The notion of stability remains the same as in the classical theory: a small perturbation (e.g. local measurements \cite{Beekman:2019pmi}) of the state should not radically alter the state. 


In practice, to observe if a SSB occurred or not we define an order parameter $O(x)$. The order parameter should be zero when the symmetry is not broken and should be different from zero when the symmetry is spontaneously broken. Ideally, it should take different values for different broken states and broken states close to each other\footnote{From one broken state we can get another one by applying the broken symmetry on it. Since we consider continuous symmetry group, the notion of ``broken states close to each other'' is understood in the sense of the continuous action of the symmetry group.} should correspond to close values of $O(x)$. A possible order parameter is of course the definition of SSB itself \eqref{DefSSBQMFormal}. In Quantum Field Theory it is customary to use the fundamental fields with nontrivial transformation properties under the symmetry as order parameter -- $ \phi(x)$ on-shell at the classical level and $\left\langle \phi(x) \right\rangle$ at the quantum level. We shortly designated this order parameter as the ``VEV'' (for vacuum expectation value). 

Another device than the order parameter which can provide a clue if a SSB occured is the two-point correlation function. However, we will not use this object in this lecture, hence, we will not expand on it. 

Let us finish with a brief vocabulary comment. When we speak about spontaneous symmetry breaking, the use of the term ``fundamental theory'' can be misleading. Usually in physics, the fundamental theory refers to the fundamental microscopic theory / to the fundamental UV theory. In SSB physics, the fundamental theory is the theory we have prior the spontaneous symmetry breaking, it is therefore not necessarily the UV theory. The term ``fundamental theory'' is thus used in order to contrast with the perturbation theory obtained from fluctuations around the broken state. From now on, in these lecture notes, we will refer to the fundamental theory in the sense of SSB physics.	

\begin{exercise} Show that when a symmetry group $G$ is spontaneously broken to a subset $H$, $H$ is a subgroup of $G$ \cite{Rubakov:2002fi}.\end{exercise}

\subsection{Singular limits}

The way spontaneous symmetry breaking is naively defined might suggests that it is only a pure academical concept. Indeed, we are looking for a stable solution by minimising the energy, and see if there is arbitrariness in the choice of the vacuum due to the symmetries (cf. the example of Figure \ref{MexicanHat} where there is a set of possible vacua due to the rotational symmetry). But in Nature, any physical system interacts with the outside world, at least a little. These external interactions will make such that one particular state of the system is energetically favourable. Thus, there is no more arbitrariness, no more spontaneity in the choice of the background. The background is explicitly chosen by the dynamics and so, we have an explicit breaking of the symmetries rather than a spontaneous one. 

The wondering can be deeper at quantum level where the notion of SSB might not even exists. Indeed, quantum superposition might allow the system to be in a superposition of broken states which end up as a symmetric state -- for example, the system could be in a superposition of all the classical vacua of Figure \ref{MexicanHat} (the $U(1)$-circle at the bottom of the Mexican hat), this superposition all over the $U(1)$-circle does not choose a specific direction anymore and so, rotation symmetry is re-established. The same argument can hold for thermal physics where the thermal state being an average of microscopic states, in average the thermal state will be/can be symmetric. 

The answer to these questionings is given by the singular limits which, when satisfied, ensure that quantum superposition will not systematically provide a symmetric vacuum. These limits guarantee as well that the external perturbations which explicitly breaks the symmetries are small enough to be non-observable. So, the explicit breaking is non-physical (because non-observable) and we can indeed speak about spontaneous symmetry breaking in Nature.  

\begin{exercise} Read about these singular limits. A starting point could be \cite{Beekman:2019pmi, Rubakov:2002fi}.
\label{SinglularLimiExercise}
\end{exercise}

The fact that we are working in Minkowski spacetime and so, in an infinite spatial volume, ensures the singular 
limits to be satisfied. Hence, in our framework, we evade the above-mentioned conceptual problems about SSB. 

\subsection{Mass and gap}

Goldstone's theorem refers to masssless particles. The notion of mass is a central idea of Goldstone physics. This is the reason why we will briefly remind here this standard notion. Let us notice that we work on Minkowski spacetime and therefore, we do not do general relativity or QFT on curved spacetime. Hence, we will not encounter the related difficulties to define energy and momentum as well as their conservation. The mass we are going to discuss is the QFT textbook definition.  

In classical field theory, the square of the mass is given by the coefficient of the quadratic no-derivative term in the action ($m_0^2 \, \phi \phi$). In this work we are just concerned if this term is present or not, which will tell us if the associated field is massive or not. For a quantum particle, we use the relativistic definition which is, its mass is its energy in the zero-limit of the $(d-1)$-momentum. The classical and quantum definitions are consistent. Indeed, the particle states of a QFT correspond to the asymptotic states of the theory, i.e. the spectrum we get while quantising the free theory. From standard QFT textbooks, quantising a free theory tells us that the energy corresponds to the dispersion relation, which in the free case is of the form $\omega^l = v\, p^n+m_0^2 $, where $l$ and $n$ are respectively the number of time-derivative and the number of space-derivative (of the dominant terms), and $v$ is a constant. Sending the momentum $p$ to zero, we see a correspondence between the classical mass and the quantum mass. In particular, when $l=2$ (it could be the relativistic case $l=n=2$), we recover the standard idea that the square of the mass is $m_0^2$. 

However, even if the two definitions are consistent, due to the renormalisation, the quantum mass might be different from the classical mass. The classical mass is a bare parameter which might need to be redefined through renormalisation conditions. Goldstone's theorem is valid at classical level as well as at quantum level. This suggests that the masses (which are zero) of the Goldstone modes are symmetry protected during the quantisation (modulo that no anomalies occur and that the SSB is not altered by the quantisation). A specific computation at one loop for the linear sigma model is done in \cite{Peskin:1995ev} to illustrate this assertion.    

Finally, to define the mass we need the energy and the momentum to be defined. Hence, we need continuous spacetime translation symmetries for our theory. For the explicit computations and proofs we will perform, we will assume to have such symmetries. However, it is not uncommon that physical systems do not have continuous spacetime translation symmetries. For example, crystal lattices do have ``only'' discrete spatial translations. Another example could be open macroscopic systems which do not have time translation symmetry since the external world can at any time modify the value of the conserved quantities (cf. the chemical potentials). For these kinds of examples, it is possible to generalise the notion of mass, we then speak about gaps. Even though we will compute assuming spacetime translations invariance, we will mention if the same final results can be obtained by relaxing this hypothesis.   

\section{Goldstone's theorem}

Goldstone's theorem has been mainly established in the sixties and has been refined over the following decades. It is Nambu who first conjectured the existence of a relation between symmetries and masses \cite{Nambu:1960tm, Nambu:1961tp}\footnote{Let us mention that the second cited paper is in collaboration with Jona-Lasinio. The first cited paper is written by Nambu alone and is older than the second cited one. Furthermore, Nambu being the common thread between the two papers, he is considered as the principal investigator of the conjecture.}. Goldstone improved the conjecture of Nambu by specifying the notion of spontaneous symmetry breaking and by stressing the importance that the broken symmetry should be continuous \cite{Goldstone:1961eq}. Goldstone, Salam and Weinberg provided two general poofs of the conjecture in \cite{Goldstone:1962es}. Following this publication, several other papers came out in order to clarify under which hypotheses Goldstone's theorem is valid \cite{Klein:1964ix, Gilbert:1964iy, Higgs:1964ia, Guralnik:1964eu, Lange:1965zz, Lange:1966zz, Watanabe:2011dk, Watanabe:2011ec}\footnote{Since we are in the historical genesis of Goldstone's theorem, it should be mentioned that, sometimes in the literature, NG modes are labelled as pions. This because historically, NG modes were studied in the framework of particle physics and light mesons analysis (e.g. \cite{Nambu:1961tp, Coleman:1969sm, Callan:1969sn}). For example, \cite{Leutwyler:1993iq, Leutwyler:1993gf} which are cornerstone papers in the building of effective theory for NG modes call the latter pions.}. Some alternative (formal/axiomatic) proofs and corollaries have also been provided, e.g. \cite{Bludman:1963zza,Streater:1965zja,Kastler:1966wdu,PhysRevLett.16.370}. These research efforts led to the current statement of Goldstone's theorem.

\begin{theorem}[Goldstone's theorem]
 Let us consider a physical (field) theory at the quantum level, respectively at the classical level, with a global continuous symmetry group $G$ such that it is spontaneously broken to a subgroup $H$ different from $G$ ($H \varsubsetneq G$) and that the notion of gap is well defined. Then, the spectrum of the theory will contain at least one gapless particle, respectively at least one gapless mode.
 \label{GoldstoneTheorem}
\end{theorem} 

Let us notice how generic $G$ can be: it can be uniform or non-uniform, involving spacetime symmetries or not, being compact or not etc. Furthermore, the theorem is relatively loose concerning the notion of mass (therefore we speak about gaps). The theorem is thus valid for theories defined on crystal lattice, for open systems etc. Finally, the locality requirement of the theory is hidden in the ``physical'' aspect. More explicitly, the interactions should at most have a finite range or an
exponential spatial decay (otherwise, the validity of the theorem should be checked case by case) \cite{Lange:1965zz, Lange:1966zz, Brauner:2010wm, Watanabe:2011ec}. In conclusion, Goldstone's theorem is very general!

There are two proofs of Goldstone's theorem, one which is using the quantum effective action formalism and one which is established in the Dirac notation of quantum mechanics. The gist of the first one can be understood based on the intuitive picture we will provide later. The second proof is more strict on the hypothesis than what is mentioned for Theorem \ref{GoldstoneTheorem} but it permits to display straightforwardly the spectral content. It is the proof based on the spectral decomposition of Dirac bra-ket that we will present. 

\begin{exercise} Read about the quantum effective action proof of Goldstone's theorem from the original papers \cite{Goldstone:1962es, Watanabe:2011ec} or from the textbooks \cite{Peskin:1995ev, Weinberg:1996kr,Rubakov:2002fi} (Rubakov book remains at the classical level). \end{exercise}

\subsection{Spectral decomposition proof}
\label{Spectral decomposition proof}

As already mentioned, to perform the spectral decomposition proof we have to revise some hypotheses of Theorem \ref{GoldstoneTheorem}. We will consider the group $G$ to be global continuous and uniform. The continuity of $G$ allows to define Noether currents and asking $G$ to be global permits to evade gauge symmetries for which the conserved currents (defined through Noether first theorem) are trivial (in the sense of the equivalence relation) \cite{Barnich:2018gdh, Ruzziconi:2019pzd}. As we will see, Goldstone modes rely on these conserved currents, we therefore wish to avoid any technical definitions (cf. Noether second theorem). A connected argument is through the Brout-Englert-Higgs mechanism \cite{Anderson:1958pb, Anderson:1963pc,Schwinger:1962tn,Higgs:1964pj,Englert:1964et} which illustrates that some of the NG modes are absorbed by gauge transformations and are thus unphysical. In order to ensure that Goldstone's theorem systematically leads to at least one physical massless mode, we safely chose to consider $G$ as global. The last constraint on $G$, i.e. to be uniform, is imposed to avoid the case of spacetime symmetries (spacetime translations symmetry breaking will be ruled out by considering a homogeneous vacuum) and to ease the technicalities of the computations. 

\begin{exercise} Read about the Elitzur theorem which, briefly stated, says that local symmetries cannot spontaneously be broken at quantum level \cite{Elitzur:1975im}. This is an additional argument against local $G$ for Goldstone's theorem. The intuition of Elitzur's theorem can be made through the exercise on the singular limits.  In Rubakov book \cite{Rubakov:2002fi}, it is mentioned that at quantum level, the system cannot be in a symmetric superposition of broken states because the action of a global broken symmetry to go from one vacuum to another requires an infinite energy in large volume limit. Why does this argument not hold anymore for gauge symmetries ? \end{exercise}

Now that we are ensured to have properly defined conserved currents, we have to guarantee to be able to associate to them conserved charges. This is done by specifying what we mean by locality. We will ask the interactions to be local enough such that
\begin{equation}
\int_{\partial V}dS \, j^i(x) = 0 \ ,
\end{equation}
where $V$ is the spatial volume of the system. Hence,
\begin{equation}
\frac{d  Q}{dt} = \int_{V} d^{d-1} x \, \partial_0 j^0 = - \int_{V} d^{d-1} x \, \partial_i j^i = - \int_{\partial V}dS \, j^i(x) = 0 \ .
\end{equation}
A more precise statement is made in \cite{Guralnik:1967zz}.

As cited above, we are not considering the spontaneous breaking of spacetime symmetries. It implies that our vacuum $\ket{0}$ is homogeneous (i.e. an eigenstate of $P_{\mu}$). Also, since we do not include gravity, we chose $\ket{0}$ to be the zero of energy: $P_{\mu}\ket{0}=0$. 

Finally, the main hypothesis of Goldstone's theorem is that we have spontaneous symmetry breaking. Let $Q$ be a generator of $G$ such that $Q$ is spontaneously broken. By definition, it exists a field $\Phi$ giving
\begin{equation}
\bra{0}[Q,\Phi(x)]\ket{0} \neq 0 \ .
\label{SSBdefForProof}
\end{equation}

To prove Goldstone's theorem under the aforementioned hypothesis, we study the spectral decomposition of \eqref{SSBdefForProof} by injecting a closure relation where the basis vectors $\ket{n_{\vec{k}}}$ are eigenvectors of $P_{\mu}$ \cite{Brauner:2010wm}. 

\begin{equation}
\begin{aligned}
\bra{0}[Q,\Phi(x)]\ket{0}&=\int d^{d-1}x' \bra{0}[j^0(x'),\Phi(x)]\ket{0} \ , \\
&=\int d^{d-1}x' \sum\limits_{n}\int \frac{d^{d-1}k}{(2\pi)^{d-1}}\left(\bra{0}j^0(x')\ket{n_{\vec{k}}}\Bra{n_{\vec{k}}}\Phi(x)\Ket{0}\right. \\
& \phantom{=\int d^{d-1}x' \sum\limits_{n}\int \frac{d^{d-1}k}{(2\pi)^{3}}}\left. -\bra{0}\Phi(x)\Ket{n_{-\vec{k}}}\Bra{n_{-\vec{k}}}j^0(x')\Ket{0} \right) \ . \\
\end{aligned}
\end{equation}
As we did in \eqref{translationToOriginConsCur}, thanks to $Q$ that generates a uniform symmetry, we translate the conserved current to the origin:
\begin{equation}
\begin{aligned}
\bra{0}[Q,\Phi(x)]\ket{0}&=\int d^{d-1}x' \sum\limits_{n}\int \frac{d^{d-1}k}{(2\pi)^{d-1}} \, e^{-i k_{\mu} x'^{\mu}}\left(\bra{0}j^0(0)\ket{n_{\vec{k}}}\Bra{n_{\vec{k}}}\Phi(x)\Ket{0}\right. \\
& \phantom{=\int d^{d-1}x' \sum\limits_{n}\int \frac{d^{d-1}k}{(2\pi)^{d-1}}}\left. -\bra{0}\Phi(x)\Ket{n_{-\vec{k}}}\Bra{n_{-\vec{k}}}j^0(0)\Ket{0} \right)  \ ,\\
&= \sum\limits_{n}\int d^{d-1}k \, e^{-i E_n(\vec{k}) t}\,\varphi(\vec{k})\left(\bra{0}j^0(0)\ket{n_{\vec{k}}}\Bra{n_{\vec{k}}}\Phi(x)\Ket{0}\right. \\
& \phantom{=\int d^{d-1}x' \sum\limits_{n}\int \frac{d^{d-1}k}{(2\pi)^{3}}}\left. -\bra{0}\Phi(x)\Ket{n_{-\vec{k}}}\Bra{n_{-\vec{k}}}j^0(0)\Ket{0} \right)  \ ,\\
\end{aligned}
\label{FinalExprOfSpecDecomp}
\end{equation}
where,
\begin{equation}
\int  \frac{d^{d-1}x'}{(2\pi)^{d-1}} e^{i \vec{k} \vec{x}} = \varphi(\vec{k}) \xrightarrow[V \rightarrow + \infty]{} \delta^{d-1}(\vec{k}) \ . 
\label{deltaDiracLimit}
\end{equation}
From \eqref{deltaDiracLimit}, we have that only the modes in the zero-momentum limit intervene in the integral of \eqref{FinalExprOfSpecDecomp}. Furthermore, since we have $dQ/dt =0$, it means that the only time dependence in \eqref{FinalExprOfSpecDecomp} is coming from $\Phi(x)$. Thus, the exponential should not intervene. Therefore, only the modes with 
\begin{equation}
E_n(\vec{k}) \xrightarrow[\vec{k} \rightarrow \vec{0}]{} 0 \ ,
\end{equation}
i.e. the massless modes, should contribute to the sum over $n$. Finally, with the hypothesis \eqref{SSBdefForProof}, the final result should be non-zero. Thus, we are ensured there exists at least one particle\footnote{The term ``particle'' is used in a generic way. Since these degrees of freedom are not the fundamental ones, we usually speak of quasi-particles or collective excitations.} $\ket{n_{\vec{k}}}$ which is massless and which is such that $\bra{0}j^0(0)\ket{n_{\vec{k}}}\Bra{n_{\vec{k}}}\Phi(x)\Ket{0}\neq 0 $. These are the NG modes, we learned that, besides being massless, they are created by the action of the broken symmetry on the vacuum (here represented by $j^0(0)\Ket{0}$) in a way which still needs to be clarified/formalised (cf. the coset construction). 

The reason we did not write $\varphi(\vec{k})$ directly as a Dirac delta is to emphasise that the evaluation of \eqref{FinalExprOfSpecDecomp} at $\vec{k} =0$ should be understood as a limit (coming from the infinite volume limit). Hence, we should not consider isolated momentum eigenstates with zero eigenvalue (i.e. spurious states). So, the NG modes are indeed properly defined particles. A more detailed discussion on the spurious states can be found in the literature (e.g. \cite{Nielsen:1975hm,Lange:1965zz, Lange:1966zz,Guralnik:1967zz}). From this discussion, we learn that the hypothesis on locality is not only there to guarantee charge conservation but also to avoid spurious states which would invalidate our conclusion on \eqref{FinalExprOfSpecDecomp}. The final outcome on locality is that the theory should have a well behaved range of interactions (at most finite range or exponantially decreasing with distance). If it is not the case, it should be checked case by case if the Noether charges are time independent \cite{Brauner:2010wm}. Notice that field theories with non-local interaction terms which cannot be written as a single spacetime integration could then be allowed \cite{Watanabe:2011ec}. 

To avoid computational heaviness, we looked at a homogeneous vacuum. But this hypothesis excludes a large area of applications in condensed matter. The spectral decomposition proof can be generalised such that it is valid in the case where the fundamental theory possesses continuous spatial translations symmetry in some directions and discrete ones in the remaining spatial directions and in the case of spatial translations symmetry breaking to discrete lattices \cite{Watanabe:2011dk}. 

\subsection{Intuitive picture of Goldstone's theorem}
\label{Intuitive picture of Goldstone's theorem}

Goldstone's theorem and its hypothesis can be understood intuitively. We have that the broken state is degenerated. Indeed, we can get a set of broken states by applying successively the spontaneously broken symmetries on our broken states. Let us call this set of so obtained broken states the coset space\footnote{This nomenclature can be understood by seeing the spontaneously broken symmetries as elements of $G/H$. The broken states being obtained by the successive action of the spontaneously broken symmetries, they are parametrised by the coset space $G/H$.}. If we do the shortcut that the symmetries of the theory are also the symmetries of the energy, we have that all the broken states of the coset space have the same energy. For simplicity, we do not consider the breaking of spacetime symmetries. Hence, there is no interplay between the kinetic energy and the potential energy while applying the spontaneously broken symmetries on the broken states. So, the broken states of the coset space have the same potential energy. Furthermore, the broken symmetries are continuous, which means that the coset space is continuously connected as well. Therefore, there is no potential hill between the broken states. A possible visualisation is to consider the Mexican hat potential example of Figure \ref{MexicanHat}. The degenerated broken states correspond to the $U(1)$-circle lying at the bottom of the potential\footnote{Notice that the $U(1)$-circle is indeed the coset space $G/\{e\}$, where $e$ is the identity, corresponding to the full spontaneous breaking of $U(1)$ symmetry.}, we do indeed observe that there is no potential hill between them. Thus, fluctuations around a chosen broken state in the directions of the broken symmetries will at quadratic order not have potential terms in the pertubation Lagrangian. Hence, such fluctuations are massless. These are precisely the (candidate\footnote{It remains to see if these fluctuations are independent of each other.}) NG modes! NG modes correspond to a spacetime modulated action of the spontaneously broken symmetries on the considered background. 

This schematic reasoning allowed us to understand why SSB leads to massless modes and in particular, why the continuity of $G$ is crucial to reach masslessness. Concerning the global aspect of $G$, the intuition was already commented in the previous section -- it permits to evade the Brout-Englert-Higgs mechanism and to make sure that the NG modes are observable. 

With this intuitive picture we can go even further on collecting information on the properties of the NG modes. We have that the Fourier transform of the spacetime modulated action of the spontaneously broken symmetries tell us how fast these modulations fluctuate through spacetime. If we go in the IR, usually we use the scale of the VEV to determine what low energy means, it is equivalent to look for modulations with small wave vectors $k^\mu$ and thus fluctuating with long wavelength. In the zero  $k^\mu$ limit, the modulations become constant over spacetime. So, in this limit, the modulated action of the spontaneously broken symmetries is nothing else than the regular action of symmetries joining two vacua. Hence, the NG modes do no provide additional energy to the background. This is the signature that they do not interact. We arrive at the conclusion that in the IR, the NG modes are weakly coupled. Finally, it is customary to hear about the NG modes as ``NG bosons''. This is because many of the practical cases involve only internal spontaneously broken symmetries. Indeed, the action of such symmetries does not mix the Lorentz group representations (the symmetry algebra commutes with Lorentz algebra). So to speak, the algebra of $G$ is spin zero, hence, the fluctuations produced by such elements by acting on the vacuum are scalars. The common example of NG modes with a non-trivial spin are the Goldstinos coming from the spontaneous symmetry breaking of supersymmetry which mix non-trivially the Lorentz representations -- in particular it links a boson with a fermion, we thus understand that the fluctuations should have a non-trivial spin.

We learned that the definition of an NG mode is a fluctuation around the background in the direction of one of the spontaneously broken generators. A thorough analysis is still needed to establish if the NG modes are independent or not, but we know that the related independent degrees of freedom are massless and weakly coupled in the IR. Furthermore, the spontaneous breaking of internal symmetries leads to scalar NG modes, we then qualify them as (Nambu) Goldstone bosons.

\subsection{Toy model: spontaneous symmetry breaking of $U(1)$}
\label{Toy model: spontaneous symmetry breaking of U1}

Till now, we remained abstract in the development of what Goldstone's theorem is. The aim of this subsection is to illustrate the different results we obtained so far with a concrete example. To do so, let us consider the following toy model of a complex scalar field in $d \geq 2$ dimensions
\begin{equation}
\mathcal{L}= \partial_{\mu}\phi^{*}\partial^{\mu}\phi + M^{2}\phi^{*}\phi - \lambda(\phi^{*}\phi)^{2} \;\;\;\text{ with } M^{2}>0 \text{ and } \lambda>0 \ .
\label{U(1)ToyModel}
\end{equation} 
The potential term $V(|\phi|)=- M^{2}\phi^{*}\phi + \lambda(\phi^{*}\phi)^{2}$ is the Mexican hat potential of Figure \ref{MexicanHat}. 

We can observe that the theory \eqref{U(1)ToyModel} is invariant under the $U(1)$ symmetry 
\begin{equation}
\phi(x) \rightarrow e^{i \alpha} \phi(x) \ .
\end{equation}

To find a stable background, we look for a particular solution of the EOM which minimises the energy. We ask this solution to be a non-zero constant $\phi_0$ to minimise to kinetic energy. Concerning the potential energy, we impose
\begin{equation}
\left. \frac{dV(|\phi|)}{d |\phi|}\right|_{\phi_0} = 0 \; \Leftrightarrow \; |\phi_0| = \sqrt{\frac{M^2}{2 \lambda}} \equiv v \ .
\label{PotentialMinimisation}
\end{equation}
From the energy minimisation, the phase remains unspecified (it corresponds to the $U(1)$ circle at the bottom of the Mexican hat), therefore, we will arbitrarily choose it to be zero. Notice that $\phi_0 = 0 $ would also extremise the energy but it would correspond to a maximum and thus, to an unstable background. So, our particular solution is $\phi_0(x)=v$ (it is straightforward to check that it is indeed a solution of the EOM). It breaks spontaneously $U(1)$ symmetry because it transforms non-trivially under it
\begin{equation}
v \rightarrow  v \, e^{i \alpha}  \ .
\end{equation}
We have that all the hypothesis of Goldstone's theorem are satisfied (check it). Hence, we expect to find at least one massless mode in the pertubation theory. To explicitly verify it, we parametrise the fluctuations as 
\begin{equation}
\phi(x) = (v + \sigma (x) )\, e^{i \theta(x)} \ .
\end{equation} 
The pertubation theory till third order is 
\begin{equation}
\mathcal{L}=\partial_\mu \sigma \partial^\mu \sigma + v^2 \partial_\mu \theta \partial^\mu \theta - 2 M^2 \sigma^2 - 2 \sqrt{ 2 \lambda M^2}\, \sigma^3 + 2\, v \,\sigma\, \partial_\mu \theta \partial^\mu \theta + \mathcal{O}\left( \epsilon^4 \right) \ .
\label{perturbTheoToyModel}
\end{equation}
where $\epsilon \sim \theta \sim \sigma$. We observe that $\theta(x)$ is a massless mode. Is it the predicted NG mode or is it a matter of luck ? We have that $\theta(x)$ parametrises a perturbation of the vacuum $v$ in the direction of the action of $U(1)$
\begin{equation}
v \rightarrow v \, e^{i\theta(x)} \ ,
\end{equation}
where the arrow corresponds to a spacetime modulated $U(1)$ action. We see that $\theta(x)$ is by definition an NG mode and so, the prediction of Goldstone's theorem is indeed satisfied for our toy model. 

If we pay attention to the interaction term of the NG mode $\theta(x)$, it involves derivatives. If we go in Fourier space, we understand that this interaction term will go to zero at low energy. We recover the idea that NG modes are weakly coupled in the IR. 

We could have guessed the shape of the interaction terms for $\theta(x)$ based on the $U(1)$ symmetry. Indeed, the fundamental theory \eqref{U(1)ToyModel} is invariant under $U(1)$ and so should be the perturbation theory \eqref{perturbTheoToyModel}. The transformation rule of $\theta(x)$ is given by
\begin{equation}
(v + \sigma (x) )\, e^{i \theta(x)} \rightarrow (v + \sigma (x) )\, e^{i \theta(x)} e^{i \alpha} \ ,
\end{equation}
so,
\begin{equation}
\theta(x) \rightarrow \theta(x) + \alpha \ .
\label{trsfLawNGTOyModel}
\end{equation}
We have that $\theta(x)$ transforms as a shift and $\sigma(x)$ is invariant. Therefore, the perturbation theory \eqref{perturbTheoToyModel} is $U(1)$ invariant if on each $\theta(x)$ there is a derivative acting on it. This explains why $\theta(x)$ is massless and why its interactions go to zero at low energy. 

\begin{exercise} Derive this toy model by using Mathematica and look at higher orders that indeed, $\theta(x)$ always appears with a derivative in front of it in \eqref{perturbTheoToyModel}.\end{exercise}

We can now have a sense of why we say that the symmetries which are spontaneously broken are ``hidden''. It is because they are non-linearly realised in the perturbation theory (cf. \eqref{trsfLawNGTOyModel}) which makes it not always convenient to see the symmetry invariance. In addition, NG modes are transforming non-homogeneously (cf. \eqref{trsfLawNGTOyModel}, even if we evaluate $\theta(x)$ at zero, it still transforms) which explains the systematic derivative operators acting on them. 
 
We should emphasise that the intuition we acquired from subsections \ref{Intuitive picture of Goldstone's theorem} and \ref{Toy model: spontaneous symmetry breaking of U1} is valid for the spontaneous breaking of internal symmetries. When the breaking of spacetime symmetries is involved, there are additional conceptual and technical difficulties which might spoil some of the intuitive results we derived so far (but the masslessness aspect of NG modes claimed by Goldstone's theorem remain robustly true).  

\subsection{Directions of research}

Goldstone's theorem provides an interesting observation. But, as often in science, an interesting observation leads to new questionings. Here are listed some of the main open questions brought by Goldstone's theorem:
\begin{enumerate}
\item Goldstone's theorem predicts the existence of gapless modes when a global continuous SSB occurs but does not provide a precise statement on how many there will be in the pertubation theory. We thus need a counting rule for such modes, said otherwise, we need to know which NG modes candidates are dependent and independent from each other. To do such counting, we probably also need a classification of the NG modes.
\item The fundamental hypothesis of Goldstone's theorem is that we have a SSB. Therefore, it could be meaningful to probe what are the conditions to have spontaneous symmetry breaking in a given theory. For example, Coleman's theorem \cite{Coleman:1973ci} states that, at quantum level, for relativistic theories in two dimensional spacetime, there are no SSB that could lead to NG modes. 
\item  If we know the number of NG modes and their statistics (for internal symmetries, these are bosons) it could be interesting to have their dispersion relations. This would, for example, allow us to compute thermodynamic observables. A generic study of the possible shape of dispersion relations, for instance \cite{Watanabe:2014zza}, could then be of interest. 
\item The NG modes are systematically present in the IR since they are massless. But they are not the only type of light particles. So, in the perspective of building effective field theories, we should understand how NG particles interact with other non-symmetry originated particles. An example could be the Cooper pairs where it is the interaction between the electrons and the phonons (NG modes coming from the discrete breaking of spatial translations) which display the effective attractive interaction between the electrons once we integrated out the phonons.
\end{enumerate}
Partial answers have already been provided to this list of questions but, each of these points remains an active topic of research. A brief state of the art will be made at Section \ref{Some further directions}.

\section{Counting rule through the coset construction}
\label{Counting rule through coset construction}

We already exposed the fact that Goldstone's theorem does not provide a precise statement on the number of NG modes we could expect from a given symmetry breaking pattern. It remains an open question for a totally generic breaking pattern, however, some progress and some strong results have been obtained through the past decades. We will present one of the existing counting rules and its associated classification. To derive this counting rule, we will do the hypothesis that $G$ is internal and compact (in addition to be global and continuous). We are going to establish, by using the coset construction, the most generic shape of the dominant terms in the IR of an effective field theory describing NG modes resulting from a given breaking pattern $G \rightarrow H$. Then, we will count the number of canonical independent degrees of freedom contained in this generic theory which will provide us a counting rule for the NG modes and a classification based on the broken generators. The obtained counting rule was conjectured and partially proved by Brauner and Watanabe \cite{Watanabe:2011ec} and proved by Murayama and Watanabe \cite{Watanabe:2012hr, Watanabe:2014fva} in the early decade of 2010. Their work is based on several progress in the counting and the classification of NG modes, a (non-exhaustive) list of relevant papers could be \cite{Bludman:1963zza, Nielsen:1975hm, Leutwyler:1993iq, Leutwyler:1993gf, Schafer:2001bq, Miransky:2001tw,Nambu:2004yia, Brauner:2005di, Brauner:2006xm, Brauner:2007uw, Brauner:2010wm, Watanabe:2011dk}.

\subsection{The coset construction} 
\label{Coset construction subsection} 

From a general perspective, the coset construction is the classification of the non-linear realisations of a given continuous symmetry group\footnote{More precisely, it concerns only the realisation of the elements of the group which are connected to the identity.} $G$ which reduce to linear representations when considering a continuous subgroup $\tilde{H}$ of $G$. Then, Lagrangians, which consist of an expansion in the fields and their derivatives (this defines locality), are built such that they are invariant under these specific realisations. We thus understand that the name ``coset construction'' comes from the quotient space $G/\tilde{H}$, the part of $G$ which is non-linearly realised, and from the construction method to get invariant Lagrangians. 

The coset construction was first established in the sixties in the area of elementary particles physics. Indeed, effective theories were built in order to describe light mesons. Such theories were displaying non-linear realisations of respectively the chiral group $SU(2)\times SU(2)$ and the chiral group $SU(3)\times SU(3)$ (see for example the non-exhaustive list \cite{Weinberg:1966fm, PhysRev.161.1483, Schwinger:1967tc, Wess:1967jq}). It is Coleman, Wess and Zumino who, in the end of the sixties, established a classification of all the non-linear realisations respecting the criteria defined in the preceding paragraph for a generic connected compact semi-simple internal symmetry group $G$ \cite{Coleman:1969sm}. Just afterwards, with the additional help of Callan, they set up a method which permits to build invariant local Lagrangians and to gauge the symmetry \cite{Callan:1969sn}. The coset construction is sometimes referred to CCWZ construction, the initials of the previously cited authors.  

Intuitively, if we take back our $U(1)$ Mexican hat example, we have that the Lagrangian we obtained after the spontaneous symmetry breaking reproduces non-linearly the $U(1)$ symmetry through the shift of the phase field which is nothing else than the NG mode. The obtained pertubation theory could then be likened to a particular case of the coset construction. This intuition and so, the interest of the coset construction for Goldstone physics, was formally noticed in \cite{Salam:1969rq}. They showed that effective field theories describing NG modes where corresponding to CCWZ invariant Lagrangians. From a coset construction they were able to recover all the properties of NG fields, i.e. massless modes (or with a light mass when a small explicit symmetry breaking occurs) and fields which are weakly coupled at small energy. Furthermore, by gauging the symmetries, they retrieve the Brout-Englert-Higgs mechanism. The non-linear transformations of NG fields are also discussed in \cite{Leutwyler:1993iq}.

From now on, we will consider the coset construction only in the framework of Goldstone physics. Since the transformation rules for the NG fields are settled by the transformation rules of the fundamental fields, we do not need the mathematical machinery of classifying the different equivalent transformation laws. We therefore can relax some of the hypotheses of the papers \cite{Coleman:1969sm,Callan:1969sn}. This introduction to the coset construction heavily relies on the review \cite{Burgess:1998ku}, another relevant review is \cite{Penco:2020kvy}.

\subsubsection{Hypotheses on the symmetry group}

Let $G$ be an internal continuous compact group which is faithfully linearly realised in the fundamental theory where the dimension of the realisation\footnote{We use the more general terminology ``realisation'' instead of ``representation'' since, as we will see, the action of $G$ will be non-linear for the specific parametrisation we will choose for the fields.} is finite. In such case, we can always ask \cite{Rubakov:2002fi}
\begin{align}
& \text{Tr}\left( G_\alpha G_\beta \right) = \delta_{\alpha\beta} \ , \label{metricOnG} \\
& G^{\dagger}_{\alpha}= G_\alpha \ , \label{HermitityOnG}
\end{align}
where $\{G_\alpha\}$ is the realisation of the generators of $G$. We will do a misnomer by denoting $G_\alpha$ as a generator of $G$. The choice made such that \eqref{metricOnG} and \eqref{HermitityOnG} are satisfied implies that the structure constants of the algebra are fully anti-symmetric:
\begin{equation}
[G_\alpha , G_\beta] = i f_{\alpha \beta}^{\phantom{\alpha\beta}\gamma} G_\gamma \ ,
\end{equation}
where $f_{\alpha \beta}^{\phantom{\alpha \beta}\gamma}$ is anti-symmetric in its 3 indices.

\subsubsection{Comment on the algebra}

We will consider that $G$ is spontaneously broken to a continuous subgroup $H$. Let us call $X_a$ the broken generators of $G$, and $T_A$ the unbroken ones. Since $H$ is a subgroup, we have
\begin{equation}
[T_A,T_B]= i f_{A B}^{\phantom{AB}C} T_C \;\; \Leftrightarrow \;\; \forall a, \; f_{A B}^{\phantom{A B}a}=0  \ . 
\label{HsubgroupStrConst}
\end{equation}
By using the full anti-symmetry of the structure constants and \eqref{HsubgroupStrConst}, we have
\begin{equation}
\forall a, \; f_{A a}^{\phantom{AB}B}=0   \;\; \Leftrightarrow  \;\; [T_A,X_a]= i f_{Aa}^{\phantom{Aa}b} X_b \ . 
\label{XaRepOfH}
\end{equation}
We observe that $\{X_a\}$ is a representation of $H$ (more precisely, the space generated by $\{i X_a\}$ is a representation of the Lie algebra of $H$). Let us emphasize that this last observation is always true for compact groups. This is one of the main reason why we take $G$ as being compact in our hypotheses.

\subsubsection{Coset construction for NG modes}

A fundamental field $\phi$ can be parametrised as \cite{Burgess:1998ku}
\begin{equation}
\phi(x)=U\! \left(\pi(x)\right)\chi(x) \ ,
\label{parametrisationFundField}
\end{equation}
with
\begin{equation}
U\! \left(\pi(x)\right) \equiv e^{i \pi^a(x) X_a} \ ,
\label{CosetParamDef}
\end{equation}
where $\pi^a(x)$ and $\chi(x)$ are general functions. If we particularise to $\chi(x)=v$ with $v$ being the constant VEV (no spacetime SSB), we have that 
\begin{equation}
e^{i \pi^a(x) X_a}v
\label{ParticularisationArounTheVacuum}
\end{equation}
corresponds to a spacetime modulated fluctuation around the VEV in a spontaneously broken direction of $G$. It is, by definition, a NG mode. The NG modes are therefore naturally parametrised by $\pi^a(x)$, i.e. the coordinates of a mapping between spacetime and the connected patch to the identity\footnote{This is mathematically quickly said, it should need a more formal description. But, the gist of what this mapping is geometrically is enough for our purpose.} of $G/H$. Henceforth, we refer to $\pi^a$ as the NG modes and we consider them as small pertubation fields. The Equation \eqref{CosetParamDef} is called the coset parametrisation.

To get the transformation rules of $\pi^a$, we use the realisation of $G$ on the fundamental field which leads us to 
\begin{equation}
g \, U\! \left(\pi(x)\right)\chi(x)=U\! \left(\tilde{\pi}(x)\right)\tilde{\chi}(x) \ ,
\end{equation}
where the tildes refer to the transformed fields and $g\in G$ is a shortcut to denote the realisation of $g$. We will keep using this misuse as long as it leads to no ambiguity. Since $g U\! \left(\pi(x)\right) \in G$ by associativity of the product and because 
\begin{equation}
\forall \alpha^a, \; e^{i \alpha^{\beta}G_\beta}= e^{i f^a(\alpha)X_a} \, e^{i g^A(\alpha)T_A} \ ,
\label{gAsAProductOfBrokenUnbroken}
\end{equation}
we can write
\begin{equation}
g U\! \left(\pi(x)\right) =U\!\left(\tilde{\pi}(x)\right) e^{i u^A(\pi,g)T_A} \ .
\end{equation}

\begin{exercise} Give an infinitesimal proof of \eqref{gAsAProductOfBrokenUnbroken}.\end{exercise}

\noindent
The transformation rules $\pi \rightarrow \tilde{\pi}$, $\chi \rightarrow \tilde{\chi}$ therefore satisfy
\begin{align}
& g U\! \left(\pi(x)\right) =U\!\left(\tilde{\pi}(x)\right) e^{i u^A(\pi,g)T_A} \ , \label{TrsfThetaGen} \\
& \tilde{\chi}(x)= e^{i u^A(\pi,g)T_A}\chi(x) \ . \label{TrsfChiGen}
\end{align}

\begin{exercise} Check that it is indeed a good realisation (i.e. the product of the realisation is the realisation of the product).\end{exercise}

\noindent
In general, these transformation rules are non-linear through the dependency of $u^A$ in $\pi$. The transformation law for $\pi^a$ is complicated, however, the one for $\chi$ is technically easier. This because it is a (non-linear) covariant transformation under $G$ built on a covariant realisation of $H$.

If we particularise the transformation laws \eqref{TrsfThetaGen}, \eqref{TrsfChiGen} for $g=h\in H$ and that we use \eqref{XaRepOfH}, we obtain
\begin{align}
& h U\! \left(\pi(x)\right) =U\!\left(\tilde{\pi}(x)\right) h \Leftrightarrow \tilde{\pi}^a(x) X_a = h\, \pi^a(x)  X_a \, h^{-1} \ , \label{TrsfThetaH} \\
& \chi(x)= h\, \tilde{\chi}(x) \ . \label{TrsfChiH}
\end{align}
We observe that our fields transform linearly under $H$ while generically, according to \eqref{TrsfThetaGen} and \eqref{TrsfChiGen}, they transform non-linearly under $G$. Thus, Nambu-Goldstone modes and their effective field theories could indeed be well described by the coset construction formalism where $\tilde{H}$ corresponds to $H$ (concerning internal symmetries at least). We remind that $\tilde{H}$ is defined as the continuous subgroup of $G$ which is linearly realised.

It could be instructive to develop a bit further the transformation rule of $\pi^a$ when $g$ is generated by broken generators. Infinitesimally, the left-hand side of \eqref{TrsfThetaGen} is
\begin{align}
g U\! \left(\pi(x)\right) &= e^{i \omega^a X_a}e^{i \pi^b(x) X_b} \ ,\\
& = (1+i \omega^a X_a + \mathcal{O}(\omega^2))(1+i\pi^b(x) X_b + \mathcal{O}(\pi^2) ) \ , \\
& = 1+i ( \pi^a(x)+\omega^a ) X_a  + \mathcal{O}(\epsilon^2) \ , \label{DevPitildeLeft}
\end{align}
where $\pi \sim \omega \sim \epsilon$. It should be equal to the right-hand side of \eqref{TrsfThetaGen}
\begin{align}
U\!\left(\tilde{\pi}(x)\right) e^{i u^A(\pi,g)T_A} &= (1+i \tilde{\pi}^a(x) X_a + \mathcal{O}(\tilde{\pi}^2))(1+i u^A(\pi,g) T_A + \mathcal{O}(u^2) ) \ , \\
&= 1+i u^A(\pi,g) T_A + i \tilde{\pi}^a(x) X_a + \mathcal{O}(\epsilon^2) \ , \label{DevPitildeRight}
\end{align}
where $\tilde{\pi} \sim u^A \sim \epsilon$. By comparing \eqref{DevPitildeLeft} to \eqref{DevPitildeRight}, we get
\begin{align}
\tilde{\pi}^a(x)=\pi^a(x) + \omega^a + \mathcal{O}(\epsilon^2) \ ,
\;\; u^A(\pi,g) =  \mathcal{O}(\epsilon^2) \ .
\label{trsfPiUnderCoset}
\end{align}
The transformation of the NG modes is inhomogenous. This is a signature that in the EFTs describing NG modes, there cannot be any mass terms for $\pi^a$ and the interacting terms involving NG modes should contain derivatives (i.e. weakly coupled in the IR). The coset construction allows us to recover all the characteristic features predicted by Goldstone's theorem.

Building an invariant Lagrangian directly using the transformation law of $\pi^a$ under $G$ is an involved process. We need an object which transforms covariantly with respect to $G$ and will therefore, constitute our building block. This object is obtained through the Maurer-Cartan 1-form: $dx^{\mu}U(\pi)^{-1}\partial_{\mu}U(\pi)$. It is a 1-form which takes its values in the Lie algebra of $G$. We can thus write:
\begin{equation}
U(\pi)^{-1}\partial_{\mu}U(\pi) = -i \, \mathcal{A}_{\mu}^{A}(\pi)\, T_A + i\, e_{\mu}^a(\pi)\, X_a \ .
\label{DefofeAndA}
\end{equation}
With $\gamma =  e^{i u^A(\pi,g)T_A}$ and 
\begin{equation}
U\!\left(\tilde{\pi}(x)\right) =g U\! \left(\pi(x)\right) \gamma^{-1} \ , 
\end{equation}
coming from \eqref{TrsfThetaGen}, we have 
\begin{equation}
U(\tilde{\pi})^{-1}\partial_{\mu}U(\tilde{\pi}) = \gamma \, U(\pi)^{-1}\partial_{\mu}U(\pi) \gamma^{-1} + \gamma \, \partial_\mu \gamma^{-1} \ .
\end{equation}
Hence, by recalling \eqref{HsubgroupStrConst} and \eqref{XaRepOfH}, the transformation rules are given by:
\begin{align}
& \, e_{\mu}^a(\tilde{\pi})\, X_a = \gamma \left( \, e_{\mu}^a(\pi)\, X_a  \right) \gamma^{-1} \ , \label{trsfeunderG} \\
& -i \, \mathcal{A}_{\mu}^A(\tilde{\pi})\, T_A = \gamma \left( -i \, \mathcal{A}_{\mu}^{A}(\pi)\, T_A \right) \gamma^{-1} + \gamma \, \partial_\mu  \gamma^{-1} \ . \label{trsfAunderG}
\end{align} 
We thus have that $e_{\mu}(\pi) \equiv e_{\mu}^{a}(\pi) X_a $ transforms in a covariant way under $G$ from a realisation of $H$. Therefore, $e_{\mu}(\pi)$ is the building block we were looking for. Indeed, any function of $e_{\mu}(\pi)$ which is covariantly invariant under $H$ will automatically be invariant under $G$. However, the transformation laws are spacetime dependent through their dependency in the fields $\pi_a(x)$\footnote{Let us emphasize/remind that we are looking to a global realisation of $G$, $G$ is not gauged.}. So, $\partial_\nu  e_{\mu}(\pi)$ does not transform covariantly. We need to define a good differential operator. If we look at \eqref{trsfAunderG}, we observe that $\mathcal{A}_{\mu}^A(\pi)$ transforms as a gauge field and can then be used to define a covariant derivative for $e_{\mu}(\pi)$:
\begin{equation}
\left(D_\mu e_\nu \right)^a
\equiv \partial_\mu e_{\nu}^a+ f^a_{\phantom{a}Bc}\, \mathcal{A}_{\mu}^B \, e_{\nu}^c \ .
\label{covDerivCosetConstr}
\end{equation}

\begin{exercise} With an infinitesimal expansion of $\gamma =  e^{i u^A(\pi,g)T_A}$ in \eqref{trsfeunderG} and \eqref{trsfAunderG}, find $\delta e_\mu^a \equiv e_\mu^a(\tilde{\pi}) - e_\mu^a(\pi)$ and $\delta A_\mu^A$. Show that $\left(D_\mu e_\nu \right)^a$ transforms as $e_\mu^a$. By doing so, we can be convinced that \eqref{covDerivCosetConstr} is indeed a covariant derivative. \end{exercise}

Before commenting on the construction of an invariant Lagrangian, we can try to have a sense on the way $e_{\mu}(\pi)$ and  $\mathcal{A}_{\mu}^{A}(\pi)$ depend on $\pi^a$. We have that
\begin{align}
U(\pi)^{-1}\partial_{\mu}U(\pi) &= e^{-i \pi^a X_a} \partial_\mu \, e^{i \pi^b X_b} \ , \\
& = i \, \partial_\mu \pi^b \, \left( e^{-i \pi^a X_a} X_b \, e^{i \pi^c X_c} \right) \ .
\label{DepofeAndAInPi}
\end{align}
We can notice the global $\partial_\mu \pi^b$ factor, and by comparing the obtained expression with \eqref{DefofeAndA}, we can conclude that $e_{\mu}(\pi)$ and  $\mathcal{A}_{\mu}^{A}(\pi)$ will systematically contain a derivative of $\pi^a$.

\begin{exercise} Keep developing \eqref{DepofeAndAInPi} to get the infinitesimal expressions 
\begin{align}
& e^a_\mu(\pi) = \partial_\mu \pi^a - \frac{1}{2}\partial_\mu \pi^b \pi^c f_{bc}^{\phantom{bc}a} + \mathcal{O}(\pi^2) \ , \\
& \mathcal{A}_{\mu}^{A}(\pi) = \frac{1}{2} \partial_\mu \pi^a \pi^b f_{ab}^{\phantom{ab}A} + \mathcal{O}(\pi^2) \ .
\end{align}
\end{exercise}

We can build a $G$ invariant Lagrangian by using $e_{\mu}$ and $D_\mu e_\nu$ by asking this Lagrangian to be $H$ covariantly invariant:
\begin{equation}
\mathcal{L}(e_{\mu},D_\mu e_\nu,\ldots) \text{ such that } \mathcal{L}(h\,e_{\mu}\,h^{-1},h\,D_\mu e_\nu\, h^{-1},\ldots)=\mathcal{L}(e_{\mu},D_\mu e_\nu,\ldots)  \ ,
\end{equation}
where the ellipses denote higher covariant derivatives. Let us mention that we could also add other fields than the NG modes through $\chi(x)$ which transforms covariantly \eqref{TrsfChiGen} (the derivative operator would then be given by $D_\mu \chi(x)$).

\subsubsection{Effective field theories for NG modes}

Said crudely, an effective theory is the most general theory respecting some given symmetry constrains which can be written as an expansion in energy. 

So, to establish the effective theory for NG modes associated to a given symmetry breaking pattern, we need to show that the most general $G$ invariant Lagrangian for $\pi^a$ built with $e_{\mu}(\pi)$ is equivalent to the most general $G$ invariant Lagrangian directly constructed with $\pi^a$. This is proven in \cite{Burgess:1998ku}, we will not repeat the proof and accept the statement (but it is a good exercise to read the proof). 

Furthermore, since a generic construction based on $e_{\mu}(\pi)$ is indeed totally general, we have that each term of the Lagrangian contains a derivative of $\pi^a$ (cf. the paragraph below \eqref{DepofeAndAInPi}). The expansion in energy is therefore rather natural. We are thus able to write a generic EFT and to guess the first dominant terms. In fact, we know that the EFT will systematically have a derivative in each term, the dominant terms will be the ones with the minimum number of derivatives.

For the relativistic case, since we need to contract the Lorentz indices, the minimum number of derivatives we can have is two. Hence, 
\begin{equation}
\mathcal{L}\left( \pi \right) = \frac{1}{2}g_{ab}(\pi)\partial_\mu \pi^a \partial^\mu \pi^b + \mathcal{O}(\partial^4) \ ,
\label{relLagCosetConstr}
\end{equation}
where $g_{ab}(\pi)$ is a symmetric matrix. To ensure the kinetic energy to be positive, we take $g_{ab}(\pi)$ 
to be positive definite for all $\pi^a$. The additional constrains we have to impose on $g_{ab}(\pi)$ in order for $\mathcal{L}\left( \pi \right)$ to be $G$ invariant will give us a geometric interpretation. 

Let us see the $G$ transformation 
\begin{equation}
\tilde{\pi}^a = \pi^a + \xi^a(\pi) \ ,
\label{GDiffeomor}
\end{equation}
as a diffeomorphism on $G/H$, where $\xi^a(\pi)$ generically depends on $\pi^a$. The latter statement can be seen by considering the higher terms in \eqref{trsfPiUnderCoset}. We get
\begin{align}
\mathcal{L}\left( \tilde{\pi} \right)&= \frac{1}{2}g_{ab}(\pi + \xi)\,\partial_\mu (\pi^a + \xi^a) \, \partial^\mu (\pi^b + \xi^b) + \mathcal{O}(\partial^4) \ , \\
&=  \mathcal{L}\left( \pi \right) + \frac{1}{2} \left( \xi^a \, \partial_a g_{bc} + g_{ac}\, \partial_b \xi^a  + g_{ba}\, \partial_c \xi^a \right) \, \partial_\mu \pi^b \partial^\mu \pi^c +  \mathcal{O}(\xi^2) \ ,
\end{align}
where we Taylor expanded $g_{ab}(\pi + \xi)$. Imposing $ \mathcal{L}\left( \tilde{\pi} \right)= \mathcal{L}\left( \pi \right) $ to have the $G$ invariance is equivalent to ask 
\begin{equation}
\xi^a \, \partial_a g_{bc} + g_{ac}\, \partial_b \xi^a  + g_{ba}\, \partial_c \xi^a  = 0 \ .
\label{TrsfMetricCoset}
\end{equation}
By seeing $\pi^a$ as coordinates on $G/H$ and its transformation under $G$ as a diffeomorphism, we can interpret $g_{ab}(\pi)$ as a metric with the isometry group $G$.  Indeed, \eqref{TrsfMetricCoset} is the Lie derivative with respect to $\xi^a(\pi)$, thus the constraint of being $G$ invariant is given by 
\begin{equation}
\mathcal{L}_\xi g =0 \ .
\end{equation}
This provides an interesting geometric picture because it shows that finding the dominant term of the most general relativistic $G$ invariant EFT is equivalent to looking for the most generic positive definite $G$ invariant metric on $G/H$. 

The geometrical interpretation of the coset construction has been established by \cite{Isham:1971dv}, this in the goal to make a link with general relativity. Volkov enriched this geometrical approach in \cite{Volkov:1973vd}.

\begin{exercise} It can be shown that when $\{ X_a \}$ forms a completely reducible representation of $H$ (we have already seen that it is a representation, cf. \eqref{XaRepOfH}), the set of positive definite $G$ invariant metrics on $G/H$ is parametrised by $n$ positive parameters, where $n$ is the number of irreducible representations of $H$ in $\{ X_a \}$. In particular, when $\{ X_a \}$ is an irreducible representation of $H$, the metric is unique up to a positive normalisation factor. Be convinced by this statement by reading the relevant sections of \cite{Burgess:1998ku} -- which is the main reference for this section on the coset construction.
\label{exerciseGenMetricParamByNumberIrrRep}
\end{exercise}

For the non-relativistic case, we are going to keep the spatial rotation symmetry. This leads us to 
\begin{equation}
\mathcal{L}\left( \pi \right) =c_a(\pi) \partial_t \pi^a + \frac{1}{2}g_{ab}(\pi)\partial_t \pi^a \partial_t \pi^b - \frac{1}{2}\bar{g}_{ab}(\pi)\partial_i \pi^a \partial_i \pi^b + \mathcal{O}(\partial_t^3,\partial_t\partial_i^2,\partial_i^4 ) \ ,
\label{nonRelLagCosetConstr}
\end{equation}
where $g_{ab}(\pi)$ and $\bar{g}_{ab}(\pi)$ are generic positive definite $G$ invariant metrics on $G/H$ (they are proportional to each other when $\{ X_a \}$ is an irreducible representation of $H$). To have an invariant theory, we need the function $c_a(\pi)$ to be a generic covector field on $G/H$ which transforms under the $G$ diffeomorphism \eqref{GDiffeomor} as a global ($\pi$-)derivative
\begin{equation} 
\mathcal{L}_\xi c_a = \partial_a \Omega_\xi (\pi) \ ,
\label{ConstrOnCa}
\end{equation}
where $\Omega_\xi (\pi)$ is an unconstrained function on $G/H$. Indeed,
\begin{align}
c_a(\tilde{\pi}) \partial_t \tilde{\pi}^a  &= (c_a(\pi)  + \partial_b c_a \xi^b ) (\partial_t \pi^a + \partial_b \xi^a \partial_t \pi^b ) + \mathcal{O}(\xi^2) \ , \\
&=c_a(\pi) \partial_t \pi^a  + ( \xi^b \partial_b c_a + c_b \partial_a \xi^b) \partial_t \pi^a + \mathcal{O}(\xi^2) \ , \\
&= c_a(\pi) \partial_t \pi^a  + \partial_t \Omega_\xi+ \mathcal{O}(\xi^2) \ , 
\end{align}
where we Taylor expended till the first order and where we used the definition of the Lie derivative acting on a covector. We notice that under the constrain \eqref{ConstrOnCa}, the Lagrangian transforms up to a global derivative. Our goal is to have an invariant theory rather than an invariant Lagrangian. So, this transformation behaviour is tolerated for our EFT. 

Let us mention that, in the context of effective theories for NG modes, terms transforming up to a global derivative are called Wess-Zumino-Witten terms \cite{Penco:2020kvy}. To be totally generic, the classification of such terms should be added to our discussion. This classification is outside the scope of this lecture. However, a discussion of such terms for our particular case can be found in \cite{Watanabe:2014fva} which ensures that \eqref{nonRelLagCosetConstr} is indeed totally general. Some possible directions to look at to learn about these Wess-Zumino-Witten terms are \cite{Wess:1971yu, Witten:1983tw, DHoker:1994rdl, DHoker:1995fsj, DHoker:1995mfi}. 

\subsubsection{Recap on the hypotheses made for the EFT}

We established that \eqref{nonRelLagCosetConstr} corresponds to the most general dominant terms of an EFT describing NG modes in the IR -- let us notice that the limit of low energy is consistent with the massless aspect of NG modes. Some of the hypotheses were explicitly stated and others have been implicitly considered. We thus here provide a recap under which hypotheses \eqref{nonRelLagCosetConstr} has been obtained. 
\begin{itemize}
\item The spontaneous symmetry breaking pattern $G \rightarrow H$ is such that $G$ is a continuous global compact internal group with a faithful finite dimensional linear representation on the fundamental theory. And $H$ is either a continuous subgroup of $G$ or the trivial subgroup containing only the identity. The constrains on $G$ might look extensively restrictive but they are satisfied in many physical cases. Indeed, $U(1)$, $SO(3)$, $SU(N)$ are for example common groups encountered in physics where they are realised through faithful matrix representations. Hence, they do satisfy the requirements on $G$. Let us remind that we took $G$ to be compact mainly to ensure that $\{X_a\}$ is a representation of $H$. A specificity that we notably used when displaying the transformation rules. It suggests that we might relax the hypothesis on the compactness of $G$ on the condition that \eqref{XaRepOfH} is still satisfied.
   
\item We considered that NG modes are the only massless modes (or light modes) in the spectrum. So to speak, we went to enough low energy such that all the massive modes have been integrated out. Of course, the discussion remains to be generalised to the case of additional massless perturbations and/or very light modes. 

\item We made the hypothesis of continuous spacetime translation symmetries. In particular, the notion of mass is well defined. Let us mention that lattice physics could be discussed through our EFT in the continuum limit which is consistent with the IR limit. 

\item The locality of the theory has been taken such that only fields at the same spacetime position interact (modulo the infinitesimal difference due to the finite number of derivatives).

\item The dimension of spacetime $d$ has been chosen to be strictly greater than two. Indeed, we took in consideration spatial rotation symmetries, spatial rotations require at least two spatial directions. If we would have studied the case $d=2$, additional terms as $\partial_t \pi \partial_x \pi$, etc. would have been tolerated. It is discussed in \cite{Watanabe:2014fva}. Furthermore, the case $d=2$ is singular (mainly for the relativistic case) due to Coleman's theorem and requires specific attention.  

\item There might be possible additional hypotheses that we overlooked due to some lack in the mathematical rigour of the development. Since we do not claim to have an axiomatic approach of physics, we will consider these hypotheses to be encompassed in the term ``physical theory''.    
\end{itemize}

Let us mention that, when writing a general EFT, we have to keep in mind that this EFT should be consistent with a possible UV completion (as discussed in Section \ref{SettingtheformalismSection}). This could for example impose some constraints on the parameters of the theory, to avoid superluminal speeds for example.

\subsection{Counting rule and classification}

All the NG modes in \eqref{nonRelLagCosetConstr} are not necessarily dynamically independent. In fact, canonical conjugation between $\pi^a$ fields can appear if the Lagrangian is of the form
\begin{equation}
\mathcal{L} \sim \pi^1 \partial_t \pi^2 - \pi^2 \partial_t \pi^1 + \ldots \ .
\label{schematicReasoningCountingRule}
\end{equation} 
By using the definition of the canonical momentum
\begin{equation}
P_a \equiv \frac{\partial \mathcal{L}}{\partial (\partial_t \pi^a)} \ ,
\end{equation}
we find that the canonically conjugated pairs are $(\pi^1,-\pi^2)$ and $(\pi^2,\pi^1)$. Hence, we have that $\pi^1$ and $\pi^2$ are canonically conjugated and both of them form one degree of freedom\footnote{We define the degrees of freedom as the quantities for which we have set their instantaneous speed and their values at initial time to unequivocally fix the dynamics.}. From this schematic reasoning, we understand that the term $c_a(\pi) \partial_t \pi^a$ in \eqref{nonRelLagCosetConstr} is crucial for our counting rule.

To count the number of independent NG modes, we have to label which modes are conjugated to which other modes. To do so, we need a classification. A natural classification could be based on the broken generators since, as we have seen in the coset construction, it is them which generate the NG modes. Hence, we need to establish a link between the $Q_a$'s (the conserved charges which after quantisation correspond to the broken generators) and the $c_a(\pi)$'s. This is done thanks to Noether currents.

The guidelines of this subsection will be to compute the Noether currents. It will provide us a non-explicit relation between the conserved charges and the $c_a(\pi)$ coefficients. By introducing a new quantity and with the preceding relation in mind, we will be able to establish a direct contact between the $Q_a$'s and the $c_a(\pi)$'s. It will then remain to express $c_a(\pi) \partial_t \pi^a$ in terms of the $Q_a$'s in our Lagrangian. So, the canonical structure of the theory will be given in terms of the $Q_a$'s. We will thus end up with a classification and a counting rule based on the broken generators. This subsection leans on \cite{Watanabe:2011ec,Watanabe:2012hr, Watanabe:2014fva}. 

Let us start by re-expressing a bit differently the transformation law of $\pi^a$.
\begin{equation}
\pi^a(x)\xrightarrow[]{G}\pi^a(x) + \xi^a(\pi) = \pi^a(x) + \omega^\alpha h_\alpha^a(\pi) \ ,
\end{equation}
such that
\begin{equation}
h_\alpha^a(\pi)=
\left\lbrace 
\begin{matrix}
h_b^a(\pi) & \text{if }\alpha = b \\
\pi^c h_{Ac}^a& \text{if }\alpha = A
\end{matrix}
\right.
\ ,
\label{DefOfhTrsfLaw}
\end{equation}
where $h_{Ac}^a$ is constant, because $\pi^a$ transforms linearly under $H$. The action of the generators of $G$ is then given by the operators
\begin{equation}
G_\alpha = h_\alpha^a(\pi) \partial_a \ .
\label{ActionOfGenEFT}
\end{equation}

We have seen that the Lagrangian \eqref{nonRelLagCosetConstr} transforms up to a global derivative. It permits to define $K_\alpha^0$ in such manner that  $\delta\mathcal{L}(\pi)\equiv \omega^\alpha \partial_t K_\alpha^0$. In such case, the (zero component of) the Noether currents are given by (cf. \eqref{ConsCurrentWithBoundaryTerm})
\begin{equation}
j_\alpha^0(\pi) = \frac{\partial \mathcal{L}}{\partial (\partial_t \pi^a)}h_\alpha^a(\pi) -K_\alpha^0(\pi)  \ .
\end{equation} 
Let us find out the expression of $K_\alpha^0$. 
\begin{align}
\delta\mathcal{L}(\pi) &= \partial_t \Omega_\xi = \left( \mathcal{L}_\xi c_a \right) \partial_t \pi^a \ , \\
&= \left(\omega^\alpha h_\alpha^b(\pi) \partial_b c_a(\pi) + c_b(\pi) \omega^\alpha \partial_a h_\alpha^b(\pi)\right)\partial_t \pi^a  \ , \\ 
&=\omega^\alpha h_\alpha^b(\pi) \partial_b c_a(\pi) \partial_t \pi^a  + c_b(\pi) \omega^\alpha \partial_t h_\alpha^b(\pi) \ , \label{Jstep1} \\
&=\partial_t \left( c_b(\pi) \omega^\alpha h_\alpha^b(\pi) \right) + \omega^\alpha h_\alpha^b(\pi)\left(\partial_b c_a(\pi) - \partial_a c_b(\pi) \right)\partial_t \pi^a \ , \label{Jstep2}
\end{align}
where to go from \eqref{Jstep1} to \eqref{Jstep2} we used an integration by part. We know that the Lagrangian transforms up to a global derivative. So, it must exist a function $r_\alpha(\pi)$ such that 
\begin{equation}
\partial_a r_\alpha(\pi) = h_\alpha^b(\pi)\left(\partial_b c_a(\pi) - \partial_a c_b(\pi) \right) \ .
\label{FirstDefOfr}
\end{equation}
Thus
\begin{equation}
\delta\mathcal{L}(\pi)  = \omega^\alpha \partial_t \left( c_b \, h_\alpha^b + r_\alpha \right)  \ ,
\end{equation}
which gives
\begin{equation}
K_\alpha^0 = c_b(\pi)\, h_\alpha^b(\pi) + r_\alpha(\pi) \ .
\end{equation}
We are now able to compute the (zero component of) Noether currents:
\begin{align}
j_\alpha^0(\pi) & = \frac{\partial \mathcal{L}}{\partial (\partial_t \pi^a)}h_\alpha^a(\pi)  -K_\alpha^0(\pi)  \ , \\
&=  \frac{1}{2} g_{ab}(\pi)h_\alpha^a(\pi) \partial_t \pi^b - r_\alpha(\pi)\ . \label{conservedCurrentEFT}
\end{align} 
The vacuum expectation value of $j_\alpha^0$ is given by its classical vacuum value. The argument is that a renormalisation factor $j_\alpha^0 \rightarrow Z j_\alpha^0$ would spoil the commutation algebra $[j_\alpha^0,j_\beta^0]=i f_{\alpha\beta}^{\phantom{\alpha\beta}\gamma}j_\gamma^0$, \cite{Watanabe:2014fva}\footnote{An alternative approach based on the Ward-Takahashi identities recovers the similar result that a non-zero VEV of non-abelian charge densities induces a one-time derivative term in the effective Lagrangian  \cite{Leutwyler:1993gf}.}. Evaluating \eqref{conservedCurrentEFT} on the vacuum corresponds to consider the fields $\pi^a$ as vanishing fields. This can be understood from \eqref{parametrisationFundField} and \eqref{ParticularisationArounTheVacuum}. So,
\begin{equation}
\bra{0}j_\alpha^0(x)\ket{0}= \left. j_\alpha^0(x) \right|_{\pi=0}=- \left. r_\alpha\right|_{\pi=0} \ .
\label{ConservCurVac}
\end{equation}

Through equations \eqref{FirstDefOfr} and \eqref{ConservCurVac}, we have a non-explicit relation between the broken charge densities and the $c_a$'s. To make this connection more explicit, let us study the quantity
\begin{align}
\rho_{\alpha\beta} &\equiv \lim\limits_{V \rightarrow + \infty} \frac{-i}{V} \bra{0}[Q_\alpha , Q_\beta]\ket{0} \ ,
\end{align}
where $V$ is the spatial volume of our system. We can develop this expression by using the definition of a conserved charge, by translating the conserved current at the origin, by taking into account the internal aspect of $Q_\alpha$ and by considering a uniform vacuum
\begin{align}
\rho_{\alpha\beta}&= \lim\limits_{V \rightarrow + \infty} \frac{-i}{V} \int_V d^{d-1}x \,\bra{0}[Q_\alpha , j_\beta^0(x)]\ket{0} \ , \\
&= -i \bra{0}[Q_\alpha , j_\beta^0(0)]\ket{0} \ .
\end{align}
With the help of \eqref{quantisationConsCharg}, \eqref{ActionOfGenEFT} and \eqref{ConservCurVac},
\begin{align}
\rho_{\alpha\beta}&= - \left. \delta_\alpha j^0_\beta(0)\right|_{\pi=0} = - \left. h_\alpha^a(\pi) \partial_a j^0_\beta(0)\right|_{\pi=0} = \left. h_\alpha^a(\pi) \partial_a r_\beta(\pi) \right|_{\pi=0} \ .
\end{align}

Let us notice that if we particularise at $\beta=A$, we have $\left.\partial_a r_A\right|_{\pi=0} =0$ because $h_A^b(\pi=0) =0$ (cf. \eqref{FirstDefOfr}, \eqref{DefOfhTrsfLaw}). Thus,
\begin{equation}
\rho_{\alpha A}=0=-\rho_{A \alpha} \ .
\end{equation} 
This is consistent with $Q_A\ket{0} \propto \ket{0}$, since $Q_A$ is unbroken, which implies $ \rho_{\alpha A}=0=\rho_{A \alpha}$. We will therefore focus on $\rho_{ab}$. With \eqref{FirstDefOfr},
\begin{align}
& \rho_{ab}=h_a^c(0)h_b^d(0) \left. \left( \partial_d c_c - \partial_c c_d \right)\right|_{\pi=0} \ , \\
\Leftrightarrow \, & \left. \partial_{[d}c_{c]}\right|_{\pi=0} =\frac{1}{2}\rho_{ab} \left( h^{-1}(0)\right)_d^a \left( h^{-1}(0)\right)_c^b \ , \label{antiSymCintermsOfRho}
\end{align}
where the brackets $[\ldots]$ on the indices correspond to an anti-symmetrisation of these indices. The intuition that we can indeed invert $h_b^d(0)$ is that $\pi^a$ transforms inhomogeneously under the action of broken generators and so, $h_b^d(0)\neq 0$. More formally, $\{X_a\}$ forms a basis of the tangential space of $G/H$ at $\pi=0$ and these generators are faithfully realised by $\{ h_b^a(\pi) \partial_a \}$. Hence, the latter expression, evaluated at $\pi=0$, is as well a basis of the tangential space of the coset space at the identity. Thus, $h_b^a(0)$ is a full ranked matrix. 

We are now able to express the canonical structure of our theory in terms of $\rho_{ab}$. From our schematic reasoning \eqref{schematicReasoningCountingRule}, we can limit ourselves to the quadratic part of the Lagrangian to probe the canonical conjugated fields. Therefore, let us expand the first term of the Lagrangian \eqref{nonRelLagCosetConstr} till the quadratic order in $\pi$:
\begin{align}
c_{a}(\pi)\partial_t{\pi}^{a}&=(c_{a}(0)+\partial_{b}c_{a}\mid_{\pi=0}\pi^{b})\partial_t\pi^{a} \ , \\
&=(c_{a}(0)+\partial_{[b}c_{a]}\mid_{\pi=0}\pi^{b}+\partial_{\{b}c_{a\}}\mid_{\pi=0}\pi^{b})\partial_t{\pi}^{a} \ , \label{development of the first term of the lagrangian}
\end{align}
where the braces $\{\ldots\}$ on the indices denote a symmetrisation of these indices. To continue the development, we can notice that 
\begin{align}
\partial_{t}(\partial_{\{b}c_{a\}}\mid_{\pi=0}\pi^{b}\pi^{a})&=\partial_{\{b}c_{a\}}\mid_{\pi=0}\partial_t{\pi}^{b}\pi^{a}+\partial_{\{b}c_{a\}}\pi^{b}\partial_t{\pi}^{a} \ , \\
&=2\partial_{\{b}c_{a\}}\mid_{\pi=0}\pi^{b}\partial_t{\pi}^{a} \ .
\end{align}
Using the last equality in \eqref{development of the first term of the lagrangian}
\begin{equation}
c_{a}(\pi)\partial_t{\pi}^{a}=\partial_{[b}c_{a]}\mid_{\pi=0}\pi^{b}\partial_t{\pi}^{a}+\partial_{t}(c_{a}(0)\partial_t{\pi}^{a}+\dfrac{1}{2}\partial_{\{b}c_{a\}}\pi^{b}\pi^{a}) \ .
\label{firstTermLagDevStep1}
\end{equation}
We can drop the term with the total derivative because it will lead to a surface term in the expression of the action and due to the fact that we neglect the boundary effects, it will not influence the evolution of the system. The first term of \eqref{firstTermLagDevStep1} can be re-expressed with \eqref{antiSymCintermsOfRho}
\begin{align}
c_{a}(\pi)\partial_t{\pi}^{a}& =\frac{1}{2}\rho_{ab} \left( h^{-1}(0)\right)_c^a \left( h^{-1}(0)\right)_d^b  \pi^c \partial_t \pi^d   \ , \\
&= \frac{1}{2}\rho_{ab} \tilde{\pi}^a \partial_t \tilde{\pi}^b \ ,
\label{AlmostFinalExprCaDotPi}
\end{align}
where we did a field redefinition, i.e. a change of coordinate on $G/H$ induced by the full ranked matrix $h_b^a(0)$. Let us do the misnomer $\rho$ as being the matrix $\rho_{ab}$ instead of $\rho_{\alpha\beta}$. Since $\rho$ is a real and an anti-symmetric matrix, there exists an orthogonal change of basis such that (let us suppose that we work with this new basis since the beginning)
\begin{align}
\rho =
\begin{pmatrix}
   M_{1} & & & & & \\
         & \ddots & & & & \\
         &        & M_{m}& & &\\
         &        &      &0& &\\
         &        &      & &\ddots &\\
         &        &      & &       &0 
\end{pmatrix}
\text{ with } M_{i}=
\begin{pmatrix}
   0 & \lambda_{i} \\
   - \lambda_{i} &0
\end{pmatrix} 
\ ,
\label{rhoMatricFull}
\end{align}      
where $\lambda_{i} \neq 0$ for $i=1,\ldots,m$. We emphasise that 
\begin{equation}
\text{rank}(\rho)=\sum\limits_{i=1}^{m}\text{rank}(M_{i})=2m \Leftrightarrow m=\tfrac{1}{2}\text{rank}(\rho) \ .
\end{equation}
From \eqref{AlmostFinalExprCaDotPi} and \eqref{rhoMatricFull}, we have
\begin{equation}
c_{a}(\pi)\partial_t{\pi}^{a}=\sum\limits_{i=1}^{m}\frac{1}{2}\lambda_{i}(\tilde{\pi}^{2i}\partial_t{\tilde{\pi}}^{2i-1}-\tilde{\pi}^{2i-1}\partial_t{\tilde{\pi}}^{2i}) \ .
\label{expression of ca in term of degree of freedom alpha}
\end{equation} 
By comparison with our schematic reasoning \eqref{schematicReasoningCountingRule}, we have that the $\tilde{\pi}^{2i}$ field is canonically conjugated with the $\tilde{\pi}^{2i-1}$ field. Hence, they do form one single degree of freedom instead of two. The associated independent NG mode is called a \emph{type B} NG mode. Since $i$ is running from $1$ to $m$, we have that the number of type B NG modes, $n_\text{B}$, is given by
\begin{equation}
n_\text{B} = \tfrac{1}{2}\text{rank}(\rho) \ .
\end{equation}
Concerning the $\pi^a$ fields lying in the null part of $\rho$, cf. \eqref{rhoMatricFull}, they do not intervene in the single time derivative term of the Lagrangian and are therefore canonically independent from the other fields. Each of these $\pi^a$ represents one degree of freedom. We denote these NG modes as \emph{type A} NG modes.   

Conceptually, a type B NG mode is generated by two broken generators\footnote{Qualitatively speaking, we are going to say that two broken generators $Q_i$ and $Q_j$ are conjugated if $\bra{0}[Q_i,Q_j]\ket{0}\neq 0$. In such case, we consider that $Q_i$ and $Q_j$ generate one type B NG mode. But it remains to show that, in the chosen basis, the generators are either conjugated by pairs or are independent.} while a type A NG mode is produced by one broken generator. However, in practice, this classification might not be robust with respect to an arbitrary choice of basis in the algebra. This is discussed in \cite{Watanabe:2011ec}.  

It is interesting to notice that by looking at \eqref{nonRelLagCosetConstr} and to its Fourier transform, type B NG mode will systematically have a quadratic dispersion relation ($\omega \sim q^2$) while type A will in general have a linear dispersion relation ($\omega \sim q$). In fact, $\bar{g}_{ab}(\pi)$ might be semi-definite positive in some cases (we, here, extend a bit our hypotheses) and thus zero for some directions. In such situation, the dispersion relations are dictated by the $\mathcal{O}(\partial_i^4)$ term and so, type A NG mode would have a quadratic dispersion relation ($\omega \sim q^2$).  

With all these developments, we have recovered (with some shortcuts) the theorem established by Brauner, Murayama and Watanabe in \cite{Watanabe:2011ec,Watanabe:2012hr, Watanabe:2014fva}. This theorem can be stated as
\begin{theorem}[Brauner-Murayama-Watanabe's theorem]
Let us consider a physical field theory living in $2+1$ or above Minkowski spacetime which is invariant under translations and rotations (at least at long distances) and where no terms contain fields at two separated spacetime points (it could eventually be relaxed to an exponentially decrease of the interactions with distance). If the fundamental theory has a faithfully linearly realised global continuous internal compact symmetry group $G$ generated by $\{Q_\alpha\}$ such that it is either completely  spontaneously broken or partially spontaneously broken to a continuous subgroup $H$, this without any anomalies and explicit symmetry breaking being involved, then, considering that the associated NG modes are the only massless modes, the number of NG bosons $n_{\text{NG}}$ is related to the number of broken symmetry generators $n_{\text{BS}}$ by the equality
\begin{equation}
n_{\text{NG}} =n_{\text{BS}}- \frac{1}{2} \text{rank}(\rho) \ ,
\label{equality of Wata Brauner counting}
\end{equation}
with
\begin{equation}
\rho_{ab}\equiv\lim\limits_{V \to \infty}\frac{-i}{V}\bra{0}[Q_{a},Q_{b}]\Ket{0} \ ,
\end{equation}
where $ V $ is the spatial volume of our system in spacetime and $\{Q_a\}$ are the broken generators.
\label{BMWtheorem}
\end{theorem} 
Let us mention that the hypothesis on the locality of the fundamental theory should ensure the effective field theory to be itself local, as we required in our preceding developments.

We conclude this discussion on the Brauner-Murayama-Watanabe's counting rule with several observations and remarks. 

First, this counting rule is not totally model independent since $\rho$ is the VEV of the commutators of the broken generators of $G$. So, there is a dependency on the vacuum of the theory and on how the symmetry group is realised. Based on a thorough analysis of the topology/geometry of the coset space $G/H$ and of the presymplectic structures which can live on $G/H$, Murayama and Watanabe completely classified the possible combinations of numbers of type A and type B NG modes for a given breaking pattern $G\rightarrow H$ \cite{Watanabe:2014fva}. Hence, it is partial information on the number of NG modes which rely only on the symmetries and so, is totally model independent. 

Second, the counting rule can be sensible to the central extension of the Lie algebra at the quantum level since $\rho$ depends on the commutators of the generators acting on the Hilbert space rather than on the phase space (the projective Hilbert space). Let us mention that the central extensions of a Lie algebra $\mathfrak{g}$ are classified by the second Chevalley-Elenberg cohomology group $H^2(\mathfrak{g})$. This group is trivial for semi-simple finite-dimensional algebras, hence, such algebras do not have central extensions \cite{Lekeu:2021flo, deAzcarraga:1995jw}  

Third, we have seen that it is the one-time derivative term which canonically combined the NG modes. Since the relativistic EFT \eqref{relLagCosetConstr} does not possess this term, we can safely conclude that all the NG modes are independent and so, their number corresponds to the number of broken generators. This is in fact displayed by \eqref{equality of Wata Brauner counting}. For relativistic theories, no charged operator under Lorentz group can acquire a VEV otherwise Lorentz symmetry would be broken. Considering no central extension, we have for relativistic theories 
\begin{equation}
\rho \sim \bra{0}[Q_{a},Q_{b}]\Ket{0} = i f_{ab}^{\phantom{ab}c} \bra{0}Q_c \Ket{0} \sim \bra{0}j_c^0 \Ket{0} =0 \ ,
\end{equation} 
because $j_c^0$ is a component of the conserved current which is a Lorentz-vector \cite{Watanabe:2019xul}. Thus, for relativistic theories $\rho =0 $ and $n_{\text{NG}} =n_{\text{BS}}$ as expected. Let us emphasise that $n_{\text{NG}} =n_{\text{BS}}$ would not necessarily be true if we consider spontaneous breaking of spacetime symmetries in a relativistic fundamental theory. 

Fourth, if we force the effective theory to depend only on the independent NG modes, the obtained EFT will be complicated. Indeed, if we try to integrate out one of the two canonically conjugated fields $\pi^a$ of a type B NG mode, it will lead to non-local interaction terms in $\mathcal{L}_{\text{eff}}$ \cite{Watanabe:2014fva}. Furthermore, it will spoil the classification since a type B would then be described by one $\pi^a$ which might be interpreted as a type A. We thus understand that the considered locality is necessary for the classification to make sense.    

Fifth, if we go to higher enough energy, the two-time derivative term become dominant compared to the single time derivative term. The canonical structure would then rather be determined by $g_{ab}(\pi)\partial_t \pi^a \partial_t \pi^b$ instead of $c_a(\pi) \partial_t \pi^a$. Since we consider $g_{ab}$ as being a positive definite metric, and so a non-degenerate metric, the two $\pi^a$ fields associated to a type B NG mode will be canonically independent. This means that each massless type B NG mode has a massive partner called an almost NG mode. The case where $g_{ab}$ is semi-positive definite, thus degenerate, is discussed in \cite{Kapustin:2012cr} where a counting rule is provided to give the number of almost NG mode we could expect. 

Sixth, it should be mentioned that the counting rule \eqref{equality of Wata Brauner counting} has been independently obtained in \cite{Hidaka:2012ym} by Mori projection operator method. Furthermore, in this article they extend the discussion to the finite temperature case. During the mid-2010 decade, \cite{Takahashi:2014vua} re-derived the counting rule \eqref{equality of Wata Brauner counting} by the Bogoliubov theory and discussed how we could take into consideration other gapless modes than the NG modes in the analysis and discussed also how we could deal with spontaneously broken spacetime symmetry breaking. 

Finally, while Brauner and Watanabe conjectured the counting rule in \cite{Watanabe:2011ec}, they provided a partial proof which requires the symmetry to be uniform instead of compact and internal. Furthermore, the proof of the counting rule relies heavily on the shape of the EFT \eqref{nonRelLagCosetConstr} and on the fact that the single time derivative is due to a non-zero VEV of a charge density. Leutwyler recovered these two ingredients thanks to an EFT building method based on the Ward Identities, where he used similar hypothesis than the ones we imposed for the coset construction except that he does not require $G$ to be compact \cite{Leutwyler:1993iq, Leutwyler:1993gf}. In addition, several examples we can find in the already cited literature (e.g. the acoustic phonon analysis in \cite{Watanabe:2012hr}) also point toward the idea that the counting rule could be extended to uniform symmetries -- mainly because the shape of \eqref{nonRelLagCosetConstr} can correspond to EFTs not necessarily coming from compact groups. Moreover, lattice systems (and the associated breaking of spatial translations and spatial rotations) could be encompassed in the discussion since we are in the IR which is consistent with a spatial continuum limit. 

\section{Concrete example : ferromagnetism}
\label{Concrete example : ferromagnetism}

The aim of this section is to illustrate how the results and the technology we introduced so far can be implemented on a specific observable physical phenomenon. We do not intend to do precise phenomenological predictions but rather to show how an analysis of the symmetries alone can already provide the behaviour of the observations, and how the coset construction can lead to crude quantitative predictions. The two main references for this part of the script are \cite{Ashcroft:1976,Burgess:1998ku}

The example we are going to look at is ferromagnetism: below a critical temperature $T_c$ (usually between $10^2$ K and $10^3$ K) certain materials acquire a spontaneous magnetisation. This magnetisation come from the magnetic moment the elementary constituents (atoms, molecules, ions, electrons,...) of the considered material can have due to their spin and their orbital momentum. At high temperature, the thermal agitation randomly orients the different magnetic momenta and so, by average, there is no global magnetisation. By decreasing the temperature and depending how the elementary constituents interact, a magnetic ordering can appear, the global alignment of each magnetic momenta can generate a spontaneous global magnetic field. 

We are going to study ferromagnet at low temperature ($T \ll T_c$), low enough such that we can do the approximation to study the microscopic theory at zero temperature in order to establish the fundamental state and the excitation spectrum. Afterwards, to get the thermodynamic quantities, we will apply the statistics on our microscopic spectrum. 

Since our purpose is mainly pedagogical, we can limit ourselves to a coarse model. As a first approximation, we can reasonably consider the electrons to be localised on their corresponding atoms. These atoms will be taken as identical and we will assume that they are placed at the sites of a 3 (spatial)-dimensional Bravais lattice. Each of them should possess a non-zero total angular momentum $\vec{S}$. It is standard practice to call this angular moment ``spin'' in reference to the original Heisenberg Hamiltonian, cf. later. We will already start from an effective theory where the Coulomb interactions combined with the Pauli exclusion provide effective interactions 
among momenta described by the Heisenberg Hamiltonian. We will consider that our system can indeed be effectively described by the Heisenberg Hamiltonian:
\begin{equation}
\hat{H}=-\sum\limits_{\textbf{R}\textbf{R'}}J_{\textbf{R}\textbf{R'}}\;\hat{\textbf{S}}_{\textbf{R}}\hat{\textbf{S}}_{\textbf{R'}} \ ,
\label{Hamiltonian of Heisenberg}
\end{equation} 
where bold letters correspond to 3-dimensional vectors, $\textbf{R}$ labelises the Bravais lattice sites and $J_{\textbf{R}\textbf{R'}}$ depends on $\textbf{R}$ and $\textbf{R'}$ only by the difference $\textbf{L}=\textbf{R}-\textbf{R'}$. We will suppose $J_{\textbf{R}\textbf{R'}}$ to decrease fast enough with $\textbf{L}$ such that our requirements on locality are satisfied. Furthermore, in the case of ferromagnetic materials: $J_{\textbf{R}\textbf{R'}}>0$ $\forall \, \textbf{R},\textbf{R'} $. 

To minimise the energy, we should maximise the scalar product in \eqref{Hamiltonian of Heisenberg}. To do so, all the spins should be aligned. The direction of the global alignment is not fixed by the energy minimisation principle, let us arbitrarily chose that all the spins align in the $x-$direction.

\subsection{Analysis based on solely the symmetries}

The interactions between spins depend on the relative orientation of the spins \eqref{Hamiltonian of Heisenberg}. Hence, a global rotation of the spins will not alter the dynamics. We thus have a global $SU(2)$ symmetry. From the coset construction and the counting rule of Section \ref{Counting rule through coset construction}, we learned that it is mainly the algebra which matters. Since $su(2)\cong so(3)$, for a better visualisation, we will consider the dynamics to have a global internal $SO(3)$ symmetry. It is indeed an internal symmetry since we rotate the spins around their attach points -- we do not rotate the crystal (the space). 

The vacuum has been established and chosen such that all the spins are aligned along the $x-$direction. We thus have a spontaneous symmetry breaking of $SO(3)$ to $SO(2)$ since the vacuum is invariant under a rotation along the $Ox$ axis but does transform under any other kind of rotations. We thus have that the generators $\hat{S}_y$ and $\hat{S}_z$ are spontaneously broken. 

Another important symmetry breaking for the coset construction is the spontaneous symmetry breaking of time reversal symmetry. Indeed, we can visualise the spin of each atom due to orbital rotation and to intrinsic rotation (the actual spin) of their constituents. By inverting the flow of time, the direction of rotation will change and so, the spins will flip. The dynamics is invariant under this flip since only the relative orientation between the spins matter. However, our vacuum will transform from an alignment along $x$ to an alignment along $-x$. 

From the breaking pattern $SO(3) \rightarrow SO(2)$, and from the hypothesis of our model, we can apply both Goldstone's theorem and the Brauner-Murayama-Watanabe's counting rule. We have that
\begin{align}
&\bra{0}[\hat{S}_j,\hat{S}_j]\ket{0} = 0 \;,\, j=y,z \ , \\
&\bra{0}[\hat{S}_y,\hat{S}_z]\ket{0} = i  \bra{0}\hat{S}_x\ket{0}= i \, n\, V S \ ,
\end{align}
where $n$ is the density of atoms, $V$ is the volume of our lattice and $S$ is the norm of the spin of one atom. So,
\begin{align}
\rho =\lim\limits_{V \to \infty}\frac{-i}{V}
\begin{pmatrix}
   0 & i \, n\, V S\\
      - i \, n\, V S   & 0  
\end{pmatrix}
= \begin{pmatrix}
   0 &  \, n\, V S\\
      - \, n\, V S   & 0  
\end{pmatrix}
\Rightarrow \text{rank}(\rho) = 2 \ .
\end{align}
The low energy excitation spectrum will contain one NG mode and it will be of type B. Indeed,
\begin{align}
& n_{\text{NG}} =n_{\text{BS}}- \frac{1}{2} \text{rank}(\rho) = 2 - 1 = 1 \ , \\
& n_{B} = \frac{1}{2} \text{rank}(\rho) = 1 \ , \\
& n_{A} = n_{\text{NG}} - n_{B} = 0 \ .
\end{align}
Since the action of $SO(3)$ does not modify the Lorentz representation of the spins, the perturbations around the vacuum in the broken directions of $SO(3)$ will be scalars. The type B NG mode will thus be a boson and because it is a type B, it will have a quadratic dispersion relation. In condensed matter literature, this excitation is either called a magnon or a spin wave (since it is a fluctuation in the spin orientation which propagates through the system).

We have all the necessary tools to get the behaviour of certain thermodynamic quantities. We work in the natural units $c=\hbar=k_B =1$, where $k_B$ is the Boltzmann constant. Let us focus on the heat capacity per unit of volume:
\begin{equation}
c(T) \equiv \frac{d \epsilon}{dT} \ ,
\end{equation}
where $\epsilon$ is the energy per unit of volume. 

The magnetic contribution to $c(T)$ is computed from
\begin{equation}
\epsilon_m = \epsilon_0 + \sum_{\textbf{q}} \omega_{\textbf{q}} \, \langle n_\textbf{q}\rangle_T \ ,
\label{energyFerro}
\end{equation} 
where $\epsilon_m$ is the magnetic energy density, $\epsilon_0 $ is the vacuum energy density and $\langle n_\textbf{q}\rangle_T $ is the average number density of magnetic NG modes of wave vector $\textbf{q}$ at temperature $T$. Because we are working at finite volume (ferromagnet materials are of finite size), the values for the wave vector are discretised due to the Born–von Karman boundary conditions 
\begin{equation}
q_i = 2\pi \frac{k_i}{L_i} \, \text{ with } k_i \in \mathbb{Z},\, i=1,\ldots,d-1 \ ,
\label{densityOfState}
\end{equation} 
where $L_i$ is the length of the system in the $i$-direction.

Since the excitation modes are bosons, $\langle n_\textbf{q}\rangle_T $ is given by the Bose-Einstein statistics:
\begin{equation}
\langle n_\textbf{q}\rangle_T  = \frac{1}{V(e^{\frac{\omega_{\textbf{q}}}{T}}-1)} \ .
\end{equation}
By going to the large volume limit, we can switch the sum for an integral in \eqref{energyFerro} and use \eqref{densityOfState} to determine the density of states in the integration measure. We roughly obtain
\begin{equation}
\epsilon_m = \epsilon_0 + \frac{1}{(2\pi)^3} \int d^3 q \frac{\omega_{\textbf{q}}}{e^{\frac{\omega_{\textbf{q}}}{T}}-1} \sim \epsilon_0 +  \int_0^{+\infty} d q \frac{q^4}{e^{\frac{q^2}{T}}-1} \sim  \epsilon_0 +  T^{5/2}\int_0^{+\infty} d x \frac{x^{3/2}}{e^{x}-1} \ ,
\label{FerroEnergy}
\end{equation}
where we used the quadratic shape of the dispersion relation, we went to spherical coordinates and we made a change of variable. The most right-hand side integral is given by the Riemann zeta function 
\begin{equation}
\int_0^{+\infty} d x \frac{x^{3/2}}{e^{x}-1} = \zeta\left(\frac{5}{2}\right)\Gamma\left(\frac{5}{2}\right) \ ,
\end{equation}
which is non-zero and finite.

Finally, we have that the magnetic contribution $c_m(T)$ to the specific heat evolves with temperature as
\begin{equation}
c_m(T) \equiv \frac{d \epsilon_m}{dT} \sim  T^{3/2} \ .
\label{FerroMagnSpecHeat}
\end{equation}
This thermal behaviour at low temperature is the standard textbook law obtained through more ``usual'' condensed matter computations \cite{alma991008996889704066}. To compare it with experiment, from the hypothesis of our model, we have to consider the specific heat as well coming from the phonons, i.e. the crystal oscillations. 

\begin{exercise} 
Enumerate the number of NG modes coming from the spontaneous symmetry breaking of continuous spatial translation symmetries to their discrete subset due to the crystal lattice itself. To do so, you have to consider the extension of Theorem \ref{BMWtheorem} for uniform symmetries, which as we have discussed, is reasonable. Such NG modes are called phonons. Guess the dispersion relations of these NG modes from their types. With a similar handy reasoning we did for the magnons, show that the contribution of the phonons to the specific heat goes as $T^3$. Let us mention that there are no additional NG modes due to the breaking of spatial rotations, this will be explained in the next section.
\end{exercise}

\begin{figure}[!ht] 
 \begin{center}
  \includegraphics[scale=0.4]{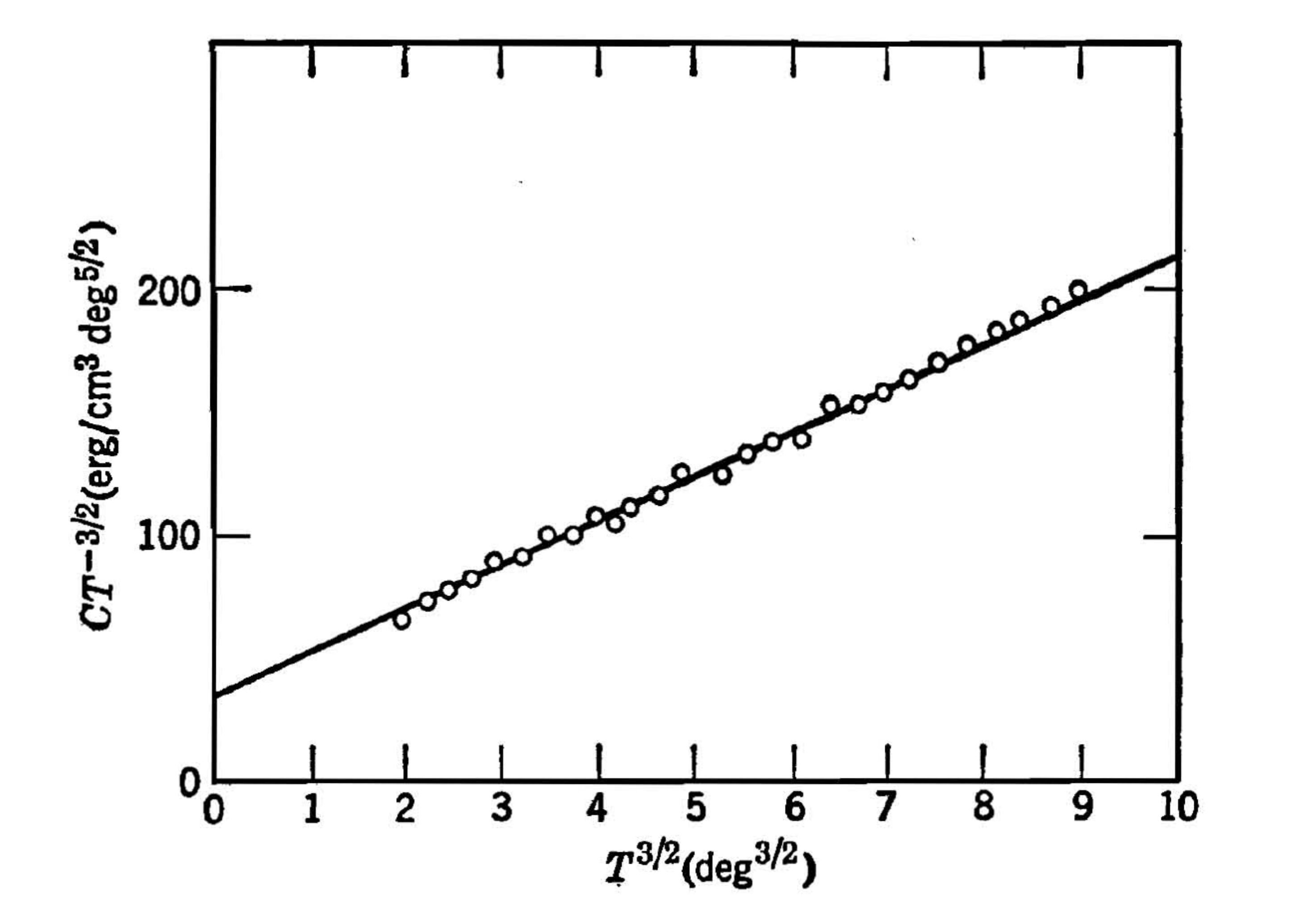}
 \end{center}
 \caption{Heat capacity of yttrium iron garnet at low temperature \cite{PhysRev.122.388, alma991008996889704066}.}
 \label{FerroSpecHeatPlot}
\end{figure}

Solely based on the symmetries involved in our model describing ferromagnets, we displayed that the specific heat receives two contributions\footnote{We made the assumption that the electrons are localised, hence, no additional electronic contribution to the specific heat is considered.} such that at low temperature, it has the following behaviour
\begin{equation}
c(T) = a \, T^{3/2} + b \, T^{3} \ ,
\end{equation} 
where the $a$ and $b$ coefficients quantify respectively the magnetic contribution and the crystal contribution. If we conduct a measurement on a ferromagnetic material with close enough properties to our models and that we plot $c(t)\, T^{-3/2}$ in function of $T^{3/2}$, we should get a straight line permitting to determine $a$ and $b$. This can be observed for Yttrium Iron Garnet from $1.5$ to $4.2$ K \cite{PhysRev.122.388}. The plot is given at Figure \ref{FerroSpecHeatPlot}.

\subsection{Coset construction for ferromagnetism}

In order to have more quantitative results for our analysis of ferromagnetism, we can build the effective field theory describing magnons through the coset construction. Indeed, the breaking pattern $SO(3)\rightarrow SO(2)$ does satisfy the criteria with which we introduced the coset construction. 

The coset space is 
\begin{equation}
G/H = SO(3)/ SO(2) \cong S^2 \ .
\end{equation}
The 2-sphere can be parametrised by the azimuthal angle $\varphi \in [0,2\pi[$ and by the angular angle $\theta \in [0,\pi]$. Thus, our candidates NG modes $\pi$ are
\begin{equation}
\pi(x) = (\theta(x),\varphi(x)) \ .
\end{equation}

Time reversal symmetry is broken, thus the EFT should contain a single time derivative term and this term will be dominant with respect to $\mathcal{O}\left(\partial_t^2\right)$ since we consider to be (deep enough) in the IR. Furthermore, the IR region is consistent with the continuum limit, and we will thus make the approximation that we have continuous spatial translation symmetries and spatial rotation symmetries. Hence, the shape of the effective Lagrangian is
\begin{equation}
\mathcal{L}\left( \pi \right) =c_a(\pi) \partial_t \pi^a - \frac{1}{2}\bar{g}_{ab}(\pi)\partial_i \pi^a \partial_i \pi^b + \mathcal{O}(\partial_t^2,\partial_t\partial_i^2,\partial_i^4 ) \ .
\end{equation}

\begin{exercise}
Show that the real vector space generated by $\{i S_y, i S_z\}$ is an irreducible representation of $so(2)$ -- a real algebra -- where the action of $so(2)$ is defined by $[\cdot , i S_x]$ \cite{Burgess:1998ku}.
\label{ExerciseIrrRepSO2}
\end{exercise} 

The coefficient $\bar{g}_{ab}(\pi)$ should be a generic metric on $S^2$ invariant under $SO(3)$. Since the broken generators form an irreducible representation of $SO(2)$ (cf. Exercise \ref{ExerciseIrrRepSO2}) and from Exercise \ref{exerciseGenMetricParamByNumberIrrRep}, $\bar{g}_{ab}(\pi)$ is given up to a global factor, a natural particular metric is the canonical metric of the 2-sphere, thus our generic metric is: 
\begin{equation}
\bar{g}_{ab}(\pi) = u_1
\begin{pmatrix}
   1& 0\\
    0   & \sin^2(\theta) 
\end{pmatrix} \ ,
\label{FerroExprMetricBar}
\end{equation}
where $u_1$ is a general constant. Another way to recover \eqref{FerroExprMetricBar} as the most generic metric is to solve the system given by the vanishing Lie derivatives of the metric with respect to the $SO(3)$ generators\footnote{
The $SO(3)$ generators realising rotations on the 2-sphere are 
\begin{equation}
\begin{aligned}
\xi^{a}_{(1)}=\delta_{\phi}^{a} \ , \;
\xi^{a}_{(2)}=-\left(cos(\phi)\delta_{\theta}^{a}-cot(\theta)sin(\phi)\delta_{\phi}^{a}\right) \ , \;
\xi^{a}_{(3)}=sin(\phi)\delta_{\theta}^{a}+cot(\theta)cos(\phi)\delta_{\phi}^{a} \ .
\end{aligned}
\label{FerroKillingVec}
\end{equation}}. 

The coefficient $c_a(\pi)$ is a generic covector which should transform as 
\begin{equation} 
\mathcal{L}_\xi c_a = \partial_a \Omega_\xi (\pi) \ ,
\label{FerroTrsfC_a}
\end{equation}
where $\Omega_\xi (\pi)$ is an unconstrained function on $S^2$ and where $\xi$ is any of the Killing vectors corresponding to the generators of $SO(3)\,$. It means that the anti-symmetric tensor
\begin{equation}
F_{ab} \equiv \partial_a c_b - \partial_b c_a \ ,
\label{FerroFabDef}
\end{equation}
is invariant under the isometry generated by $\xi$. Furthermore, since $F_{ab}$ is defined on a two-dimensional manifold (the 2-sphere), it is defined by one scalar function $u_0'(\pi)$:
\begin{equation}
F_{ab} = u_0' \sqrt{\text{Det}(\bar{g})} \, \epsilon_{ab} \ .
\label{FerroFabGenExpr} 
\end{equation}
The invariance of $F_{ab}$ reduces to 
\begin{equation}
\mathcal{L}_\xi u_0'= \xi^a \partial_a u_0' = 0 \ ,
\end{equation}
for all the rotation Killing vectors $\xi$. This condition implies that $u_0'(\pi)$ is in fact independent of $\pi$. Combining this observation with \eqref{FerroExprMetricBar}, \eqref{FerroFabDef} and \eqref{FerroFabGenExpr} we obtain an equation for $c_a(\pi)$
\begin{equation}
\partial_\theta c_\varphi - \partial_\varphi c_\theta  =  u_0' \, u_1 \sin (\theta) \ .
\label{FerroEqtForC}
\end{equation}
Because the transformation \eqref{FerroTrsfC_a} of $c_a(\pi)$ is a symmetry of the theory and it is driven by an arbitrary function $\Omega_\xi (\pi)$, we can use it to eliminate one of the components of $c_a(\pi)$. Let us set to zero $c_\theta(\pi)$. Thus, a generic solution for \eqref{FerroEqtForC} is
\begin{equation}
c(\pi) = (0, u_0 \cos (\theta)) \; \text{ with } u_0 \equiv -  u_0' u_1 \ .
\label{FerroCsol}
\end{equation}

\begin{exercise}
Explicitly check that $\mathcal{L}_\xi c_a = \partial_a \Omega_\xi (\pi)$ is satisfied for each Killing vectors \eqref{FerroKillingVec} when the solution \eqref{FerroCsol} is considered.
\end{exercise} 

Our effective Lagrangian is now of the form
\begin{equation}
\mathcal{L}\left( \pi \right) =u_0 \cos (\theta) \partial_t \varphi - \frac{u_1}{2}\left(  \partial_i\theta \partial_i \theta + \sin^2(\theta) \partial_j\varphi \partial_j \varphi \right) + \mathcal{O}(\partial_t^2,\partial_t\partial_i^2,\partial_i^4 )  \ .
\end{equation}
To connect our NG fields, $\theta(x)$ and $\varphi(x)$, to a physical interpretation, we express the 2-sphere with Cartesian coordinates
\begin{equation}
\left\lbrace
\begin{aligned}
&s_{x}=\sin(\theta)\cos(\varphi) \ ,\\
&s_{y}=\sin(\theta)\sin(\varphi) \ , \\
&s_{z}=\cos(\theta) \ ,
\end{aligned}
\right.
\label{cartesian coordinates 2-sphere}
\end{equation}
were the Cartesian coordinates correspond to the ``spin field'' (this field provides the spin we have in our material at the spacetime position $x$). The fundamental state is when all the spins are aligned in the $x-$direction, so, our spin field should be constant over space and time, and it should point toward the $x-$direction. This means that 
\begin{equation}
\left\lbrace
\begin{aligned}
&\theta_0(x)=\frac{\pi}{2} \ , \\
&\varphi_0(x)=0 \ .
\end{aligned}
\right.
\label{FerroVacuumCoset}
\end{equation}
NG modes correspond to small fluctuations around the vacuum in the broken directions, hence, to properly describe NG modes we have to infinitesimally fluctuate around \eqref{FerroVacuumCoset}
\begin{equation}
\left\lbrace
\begin{aligned}
&\theta(x)=\frac{\pi}{2} +\theta'(x)  \ , \\
&\varphi(x)=\varphi'(x) \ .
\end{aligned}
\right.
\end{equation}
Till quadratic order, we have 
\begin{equation}
\mathcal{L}\left( \pi \right) =\frac{u_0}{2}\varphi' \partial_t\theta' - \frac{u_0}{2}\theta' \partial_t \varphi'  - \frac{u_1}{2}\left(  \partial_i\theta' \partial_i \theta' + \partial_j\varphi' \partial_j \varphi' \right) + \mathcal{O}(\partial_t^2,\partial_t\partial_i^2,\partial_i^4 )  \ .
\end{equation}
From Fourier transform we can notice that we will get a unique dispersion relation 
\begin{equation}
\omega=\frac{u_1}{u_0} q^2 \ .
\end{equation}
Therefore, we recover the fact that we have a single particle corresponding to a type B NG mode. 

The two unknown coefficients $u_0$ and $u_1$ can be obtained experimentally. For example, by inelastic neutron scattering through the medium: if we know how much energy and momentum have the neutrons before and after the propagation, we can deduce how much energy and momentum were provided to the system in order to excite the modes of the solid. We thus get a curve of $\omega$ in terms of $q$. Of course, we have to discriminate the phonon excitations and the magnon excitations. The magnons are sensible to temperature, hence, by doing the experiment at various temperature we can discriminate the two kinds of excitations and extract the dispersion relation of the magnons. Once $u_0$ and $u_1$ are determined, we can explicitly compute \eqref{FerroEnergy} and obtained a value for \eqref{FerroMagnSpecHeat} or at least, an order of magnitude.

\section{Some further directions}
\label{Some further directions}

These lecture notes are an introduction to Goldstone physics and therefore, they are not meant to be exhaustive. A significant amount of the various topics we can encounter in this area of physics has been omitted to keep this work of a reasonable size but also due to the limited competence of the author.

In this last section we will briefly comeback on important well settled results that we did not mention so far. It will also permit to provide some possible further directions to look at. Hence, the interested reader will have some references from where to start if she/he wants to expand her/his knowledge in a specific subject. 

\subsection{Classification based on dispersion relations}

Prior to the counting rule based on the broken generators, Nielsen and Chadha proposed a classification relying on the dispersion relations of the NG modes \cite{Nielsen:1975hm}. This classification led to a counting rule. The guideline of the proof of the latter is the spectral decomposition of \eqref{DefSSBQMFormal}, in the same spirit as the argument we displayed for Goldstone's theorem at Subsection \ref{Spectral decomposition proof}. One of the original hypotheses of Nielsen and Chadha was that the considered fundamental theory should have continuous spatial translations symmetry. This has been relaxed in \cite{Watanabe:2011dk} such that discrete spatial translations are tolerated. 

\begin{theorem}[Nielsen and Chadha's theorem]
Let us consider a fundamental (field) theory such that the locality of the interactions implies that at quantum level, if $A(x)$ and $B(0)$ are any two local operators then
\begin{equation}
\mid\textbf{x}\mid\rightarrow\infty:\;\mid\bra{0}[A(\textbf{x},t),B(0)]\Ket{0}\mid\rightarrow e^{-\tau \mid\textbf{x}\mid},\;\tau>0 \ .
\label{ExpoDecayLocality}
\end{equation}  
If 
\begin{itemize}
\item $n_\text{BS}$ generators of the uniform symmetries of our fundamental theory are spontaneously broken (while the non-uniform ones remain untouched),
\item the notion of gap is well defined,
\item the dispersion relations of the NG modes are isotropic and can be written as a polynomial expansion at low momentum,
\end{itemize}
then we classify as type I the NG modes with an energy which is proportional to an odd power of the momentum at long wavelengths, $n_{\text{I}}$ is the number of such modes. The ones with an even power are called type II and $n_{\text{II}}$ is their number. The amount of NG particles satisfies the inequality 
\begin{equation}
n_{\text{I}}+2 n_{\text{II}} \geqslant n_\text{BS} \ .
\label{NCCountingRule}
\end{equation}
\end{theorem}

We can notice that asking the dispersion relation to be isotropic is an implicit hypothesis of rotational symmetry of the fundamental theory. Relaxing this hypothesis is discussed in \cite{Watanabe:2011dk}. The locality condition \eqref{ExpoDecayLocality} ensures an analytic behaviour of the Fourier transform and so, indirectly of the dispersion relations. Specifying that the latter should have a polynomial expansion at long wavelengths could be considered as tautological -- this tautology is not present in the original statement of Nielsen and Chadha, it is due to the reformulation made by the author of this manuscript. 

The Nielsen and Chadha counting rule is weaker than the one of Brauner, Murayama and Watanabe. However, the classification based on dispersion relations is still used in the literature since in some cases it is more practical. 

\begin{exercise}
Make a link between type A/type B NG modes and type I/type II NG modes. Notice that the counting rule of Nielsen and Chadha correctly reproduces the fact that for relativistic theories, the number of NG modes is equal to the number of broken generators (when we consider solely the breaking of internal symmetries).
\end{exercise}

\subsection{Pseudo Goldstone modes}
\label{Pseudo Goldstone modes}

In the context of spontaneous symmetry breaking, there might be similar modes to the NG modes which are present in the theory. They also are symmetry-originated and have a mass (partially) settled by the symmetries. To understand their origin, we can take back the intuitive picture we have for the NG modes. We spontaneously break the symmetries of the system by looking for a non-trivial solution which minimises the energy. Taking this solution to be homogeneous (in this section, we are mostly interested by the breaking of internal symmetries), the kinetic part of the theory does not intervene in the discussion which simplifies the reasoning. Then, the flat directions of the potential around the background correspond to opportunities to build massless on-shell fluctuations. The broken symmetries parametrise some of these flat directions and are thus, the origin of the NG modes. However, there can be additional flat directions:   
	\begin{itemize}
		\item The potential part of the action might have a bigger continuous symmetry group than the action as a whole. If these additional symmetries are spontaneously broken by the background, it leads to additional flat directions. 
		\item The equations for the potential minimisation might see some emergent symmetries which with the SSB mechanism could correspond to additional flat directions.  
		\item There might be a fine tuning among the Lagrangian parameters which makes that for a specific vacuum, the potential at this particular point has flat directions.   
	\end{itemize}
The additional symmetries coming from the potential are called approximate symmetries (since they are not exact symmetries of the full action). We understand that if we fluctuate the system around the vacuum along the flat directions associated to the approximate symmetries, we are getting massless modes \cite{Weinberg:1972fn, Georgi:1975tz}. These modes are called quasi NG modes since at classical level, they are massless but their mass is not symmetry protected when we quantise the theory (or when we follow the RG flow). The quasi NG modes represent a limitation of Brauner-Murayama-Watanabe's counting rule because it has been established considering the NG modes as being the only massless modes in the theory. A first counting rule was derived in \cite{Georgi:1975tz} but the generalization of Brauner-Murayama-Watanabe's counting rule is discussed in \cite{Takahashi:2014vua,Nitta:2014jta}. 

There is another family of modes which are closely related to the Nambu Goldstone mechanism. They appear when a symmetry is explicitly broken by a small parameter $m$ in the Lagrangian compared to the VEV. If the VEV would have spontaneously broken the symmetry then, we would have had an NG mode. If we consider the theory to have a continuous behaviour in the limit $m \rightarrow 0$, we can expect that the would-be NG mode has a small mass which goes to zero in the zero limit of $m$. This intuition was notably used by Nambu and Jona-Lasinio to describe light mesons in \cite{Nambu:1961tp}, one of the articles which led to the conjecture of Goldstone's theorem. It is Gell-Mann, Oakes and Renner who first established, still in the context of QCD, that the square of the mass of the would-be NG mode scales linearly with $m$ \cite{Gell-Mann:1968hlm}. This relation bears their name and is abbreviate as the GMOR relation. This result has been derived in several ways in the literature, mostly in the context of QCD, by using the Ward identities (e.g. \cite{Gasser:1982ap,Giusti:1998wy}) or by following the effective theory approach (e.g. \cite{Burgess:1998ku,Evans:2004ia}). In a more general context than QCD, a derivation of the GMOR relation has been done for generic relativistic effective field theories in \cite{Leutwyler:1993iq}. A proof of the GMOR law, at the level of the fundamental theory, for abelian internal symmetries in relativistic theories has been established in \cite{Argurio:2015wgr}.

We close this subsection with a clarification on the nomenclature. We call approximated symmetries, transformations which either leave the potential part of the action invariant but not the action as a whole or which are symmetries explicitly broken by a small parameter. The modes which have for origin the first case of approximate symmetries are called quasi NG modes. When the symmetry is explicitly broken by a small parameter and that the associated mode follow the GMOR relation, we call such mode pseudo NG mode. Finally, the term massive NG mode can refer either to the massive partner of a type B NG mode (i.e. an almost NG mode) either to a massive symmetry-originated mode obtained in the context of the introduction of a chemical potential (cf. next section).

\subsection{Goldstone physics at finite density}
\label{Goldstone physics at finite density}

As we have seen, spontaneous symmetry breaking occurs in several areas of physics. It could then be interesting to see how Goldstone's theorem is implemented in these different domains. With the quantum field theory formalism we employ, the implementation of the theorem in particle physics is rather straightforward. We could now look how does it apply for many-body systems at equilibrium. A first step to it is to consider QFTs still at zero temperature but with a chemical potential. This is the aim of this subsection.

For a statistical system we can associate a chemical potential to each of its conserved charges -- the microscopic dynamics is as usual dictated by either a Hamiltonian or a Lagrangian which might possesses some symmetries and therefore, some conserved quantities are defined. Switching on the chemical potential $\mu_Q$ associated to the charge $Q$ means that we consider that the external world acts on our statistical system such that it can vary the ``conserved'' charge $Q$. We are thus working in the grand canonical ensemble where $\mu_Q$ scales the energy cost when we vary $Q$. Considering our statistical system to be at equilibrium, its thermal state is dictated by the grand canonical partition function 
\begin{equation}
Z = \text{Tr}\left[ e^{-\beta \left( H - \mu_Q Q \right)} \right] \ .
\label{StatPartFunct}
\end{equation}
The trace in \eqref{StatPartFunct} represents a summation/integration on the phase space. Hence, in the zero temperature limit, i.e. $\beta \rightarrow + \infty$, we can do the saddle-point approximation. This means that the thermal state is given by the minimisation of $H - \mu_Q Q $ and by the small fluctuations around the minimum. 

From this brief recap of statistical physics, we emphasise that the microscopic dynamics is given by $H$ and that the thermal state of the system is settled by $\tilde{H}\equiv H - \mu_Q Q $. So, switching on a chemical potential means that the background, which will spontaneously break the symmetries of $H$, minimises $\tilde{H}$ instead of $H$. Thus, compared to the zero chemical potential case, the vacuum might change and also the symmetry breaking pattern $G \rightarrow H_G$ (the unbroken subgroup is now written with a sub-index to not be mistaken with the Hamiltonian). Furthermore, it is the fluctuations around $\tilde{H}$ which are physically important, thus, the gap (the mass) will be defined with respect to $\tilde{H}$  instead of $H$. 

Let us call $\ket{\mu}$ the microscopic state which minimises $\tilde{H}$. It is therefore an eigenvector of the latter operator and since we do not consider gravity, we can redefine the energy scale such that the eigenvalue is zero:
\begin{equation}
\tilde{H}\ket{\mu} = \left(H - \mu_Q Q  \right)\ket{\mu} = 0 \ .
\label{thermalMinimization}
\end{equation}
The current literature on Goldstone physics at finite density deals in three ways with the chemical potential case at zero temperature:
\begin{enumerate}
\item  We can interpret $\tilde{H}$ as generating an evolution of the system in a new time direction. Thermally speaking, it is the fluctuations evolving along this new time direction which interest us. Hence, we can effectively consider that the dynamics and the mass is given by $\tilde{H}$. Furthermore, from \eqref{thermalMinimization}, $\ket{\mu}$ does not break spontaneously time-translation with respect to the new definition of time. We see that we recover exactly the setup of Goldstone's theorem where the theory is changed from $H$ to $\tilde{H}$ and that the considered symmetry group should be the one of $\tilde{H}$. This idea is recovered in \cite{Schafer:2001bq,Miransky:2001tw}\footnote{These papers are written in the Lagrangian field theory approach. In such case, a chemical potential $\mu$ can be effectively described by gauging the symmetry to which it is related and by fixing the gauge field to be $A^{\nu}=\mu_Q \, \delta^{\nu 0}\,$ \cite{Kapusta:1981aa}. Schematically, the Lagrangian is thus of the form:
\begin{equation}
L = D_\mu \phi^* D^{\mu} \phi + \ldots =(\partial_0 + i \mu_Q) \phi^* (\partial_0 - i \mu_Q) \phi - \partial_i \phi^* \partial_i \phi +\ldots \ .
\end{equation}}.  It is an efficient way to proceed to ``get back on our feet'' and extract, without additional costs, informations on the low energy thermal excitations. Of course, since we disregard some of the physical aspects, we lose some results, as it will be confirmed later.   

\item We can tackle the problem with the standard spectral decomposition approach as we did in Subsection \ref{Spectral decomposition proof}, where the time evolution of the microscopic states (i.e. the kets) is driven by $H$ and where the gap is computed with respect to $\tilde{H}$. It is the results of this method that we present here below. It has the advantage of keeping track of the physical origin of $\mu_Q$. 

\item When $Q$ is spontaneously broken by $\ket{\mu}$, the vacuum $\ket{\mu}$ evolves non-trivially in $H$-time as it can be noticed from \eqref{thermalMinimization}. Therefore, time translation symmetry is spontaneously broken as well as maybe other spacetime symmetries (such as boosts for example). An approach based on the study of spacetime symmetry breaking can thus be used. The references \cite{Nicolis:2011pv, Nicolis:2013sga} explicitly deal with such problematic. Spacetime symmetry breaking is out of the scope of these notes, however, let us mention that the EFTs built on a generalisation of the coset construction for sapcetime symmetries permit to extract additional results compared to the ones we get with the standard spectral decomposition approach.

\end{enumerate}

Nicolis, Piazza \cite{Nicolis:2012vf} and Brauner, Murayama, Watanabe \cite{Watanabe:2013uya} showed, by a spectral decomposition of the order parameter, that the number of gapless NG modes is given by the Brauner-Murayama-Watanabe's counting rule where the considered broken generators are the broken symmetry generators of $\tilde{H}$, i.e. the broken symmetry generators of $H$ which commute with $Q$. They also showed that the remaining broken symmetry generators of $H$ lead to gapped modes where the gap is entirely fixed by $\mu_Q$ and by group theory. This can be summarised in the following theorem (NPBMW stands for the initials of the authors of \cite{Nicolis:2012vf,Watanabe:2013uya}). 
 
\begin{theorem}[NPBMW theorem]
Let us consider a system satisfying the Goldstone's theorem hypotheses where the symmetry group $G$ of the theory is restrained to be internal and compact. We switch on a chemical potential, $\mu_Q$, for a particular symmetry generated by $Q$. The thermal state of the system is driven by the free energy and the notion of gap is defined according to it. The free energy has a symmetry group $\tilde{G}$ such that $\tilde{G} \subseteq G$. The number of massless NG bosons $n_{\text{NG}}$ is related to the number of broken symmetry generators $n_{\text{BS}}$ of $\tilde{G}$ by the equality
\begin{equation}
n_{\text{NG}} =n_{\text{BS}}- \frac{1}{2} \text{rank}(\tilde{\rho}) \ ,
\end{equation}
with
\begin{equation}
\tilde{\rho}_{a,b}\equiv\lim\limits_{V \to +\infty}\frac{-i}{V}\bra{\mu}[\tilde{Q}_{a},\tilde{Q}_{b}]\ket{\mu} \ ,
\end{equation}
where $ V $ is the volume of our system in spacetime, $\ket{\mu}$ is the vacuum and $\{\tilde{Q}_{a}  \}$ are the broken generators of $\tilde{G}$.\\
The spectrum possesses some gapped modes as well, where their gap is entirely fixed by group theory and by $\mu_Q$. The number of the massive NG modes $n_{\text{mNG}}$ is given by
\begin{equation}
n_{\text{mNG}} = \frac{1}{2} \left[ \text{rank}(\rho) -  \text{rank}(\tilde{\rho})  \right] \ ,
\end{equation}
where $\rho$ is defined in a similar fashion than $\tilde{\rho}$ but using the broken generators of $G$, $\{Q_a\}$, instead of the ones of $\tilde{G}$ ($\{\tilde{Q}_{a}\}\subseteq \{Q_a\} $). Under an appropriate choice of basis for the Lie aglebra of $G$, the massive NG modes are generated by pairs of broken generators $\{Q_{\pm \sigma}\}$ ($\nsubseteq \{\tilde{Q}_{a}\}$) and their gaps are $\mu_Q\, q_{\sigma}$ where $\left[ Q,Q_{\pm \sigma} \right] = \pm q_{\sigma} Q_{\pm \sigma} \ .$
\label{NPBMW theorem}
\end{theorem}   

From an effective theory approach, Brauner, Murayama and Watanabe \cite{Watanabe:2013uya} have seen that there are additional massive NG modes for which the mass goes to zero when $\mu_Q$ is sent to zero but, this mass is different form the one of the gapped modes predicated by Theorem \ref{NPBMW theorem}. Assuming a continuous behaviour of the theory with the limit $\mu_Q \rightarrow 0$
and that the new vacuum in this limit still displays the same breaking pattern $G \rightarrow H_G$, the number of such additional massive NG modes can be obtained by counting the number of NG modes at zero chemical potential and substract the number of NG modes and massive NG modes of Theorem \ref{NPBMW theorem} at finite chemical potential. To understand the nature of these additional modes, their dispersion relations etc. a deeper analysis should be done. This is the subject of \cite{Nicolis:2013sga} based on seeing the introduction of a chemical potential as breaking the time translation symmetry. As already mentioned, we will not expand much on spacetime symmetry breaking. We can, however, already guess candidates for such additional massive NG modes. If we consider a relativistic microscopic theory at finite density (Lorentz symmetry is thus broken), the partners of the massive NG modes of Theorem \ref{NPBMW theorem} can also be massive\footnote{A massive NG mode of Theorem \ref{NPBMW theorem} corresponds here to a combination of two generators. We can therefore consider a partner which is given by the orthogonal combination of the two same generators.}. When we send $\mu_Q$ to zero, we recover our relativistic theory and so, all NG modes are type A, including our initial massive NG modes of Theorem \ref{NPBMW theorem}. Thus, their massive partners should also be massless, it makes them part of the additional massive NG modes we have when the chemical potential is switched on.

\subsection{No spontaneous symmetry breaking at low dimensions}

Spontaneous symmetry breaking is the fundamental hypothesis of Goldstone's theorem. It is therefore consistent to ask whenever a spontaneous symmetry breaking is possible. In this subsection we will enunciate some theorems which state that at lower spacetime dimensions, some spontaneous symmetry breaking patterns are impossible. In accordance with these lecture notes, we will mainly discuss the zero temperature case, then, a comment will be made for the finite temperature case. 

\subsubsection{Coleman's theorem}

From Exercise \ref{SinglularLimiExercise}, at quantum level, the quantum superposition of the possible classical vacua is avoided thanks to the large volume of spacetime: the energy required for the system to switch from one classical vacuum to another is proportional to the system volume (even if the potential directions are flat, there is still kinetic energy involved during the switching). Naively said, the lower the dimension of the spacetime is, the lower is the volume of the system. Hence, we could guess that at sufficiently low dimensions, the quantum fluctuations will be large enough to statistically give rise to a symmetric quantum vacuum. For example, if we take the $U(1)$-circle of the Mexican hat of figure \ref{MexicanHat}, the specific classically selected point of the circle playing the role of the classical vacuum will be forgot by the system due to the large quantum fluctuation around such point. Indeed, the fluctuations go all over the circle giving then a zero average state. The VEV being the order parameter, we lose the spontaneous symmetry breaking at quantum level. 

This idea has been formally stated by Coleman \cite{Coleman:1973ci} under the theorem:
\begin{theorem}[Coleman's theorem]
For relativistic physical field theories in two-dimensional spacetime, at the quantum level, there cannot be any spontaneous breaking of continuous internal global symmetries.
\end{theorem}    
The proof of Coleman \cite{Coleman:1973ci} is rather mathematical, hence, we will sketch the proof of \cite{Ma:1974tp} which is more physical and closer to the intuition we proposed earlier. It is a proof by contradiction, spontaneous symmetry breaking implies massless modes at quantum level. In two-dimensional spacetime, such massless modes induce an IR divergence which makes vanish the VEV (the order parameter) and so, we lose spontaneous symmetry breaking. The consistent picture is thus that we never have spontaneous symmetry breaking in such context.

To see this, we know that at enough low energy the relativistic Goldstone modes are described by \eqref{relLagCosetConstr} and are thus free. In such regime, the NG modes being independent from each other, we do not lose much generality by considering the specific abelian case of $U(1)$ symmetry spontaneously broken. The unique associated NG mode is denoted by $\theta(x)$. A relativistic massless free theory is a conformal theory, thus $\theta(x)$ has a propagator of the form
\begin{equation}
    			\left< \theta(x)\theta(0) \right> \; \propto \; \frac{\Gamma\left(\frac{d}{2}-1\right)}{\lvert x\rvert^{d-2}} \; \underset{d=2}{\propto} \; - \ln \left( \frac{\lvert x\rvert}{\lvert x_{0}\rvert} \right) \ .
    			\label{2pointsCorrCFT2d}
    			\end{equation} 
We observe that we have a radical change of behaviour when the spacetime dimension $d$ is equal to two. With a proper regularisation we get a logarithmic behaviour for $d=2$ where $x_{0}$ is the regulator\footnote{It should be mentioned that the logarithmic behaviour has for origin an IR divergence. Indeed, the masslessness of $\theta(x)$ means that the two-point correlator is of the form $\int d^{d-1}k (e^{ikx}/|\vec{k}|)$. In addition to the UV divergence, which is handled by a usual renormalisation, we have an IR divergence when $d=2$ which makes ill defined the associated theory. Mathematically, it is solved by introducing a mass regulator which will be sent to zero at the end. For the sake of the schematic proof, we do not display such technical details.}. Considering $\theta(x)$ as a free field, we have 
\begin{equation}
\left< \theta(x) \theta(0)\right> = \left< \theta^+\!(x) \theta^-\!(0) \right>  = \left< \left[\theta^+\!(x) ,  \theta^-\!(0)\right] \right> \ ,
\end{equation}
with $\theta \equiv \theta^+  + \theta^-$ where $\theta^+$ is associated to the positive energy modes and is proportional to an annihilation operator, $\theta^-$ is associated to the negatives energy modes and is proportional to a creation operator. We now evaluate the one-point function of the fundamental complex field $\phi$:
\begin{equation}
\begin{aligned}
\left< \phi(x) \right> & =\left< (v+\sigma(x))e^{i\theta(x)} \right> \ ,\\
&= v \left< e^{i\theta(x)}\right> =v \left< e^{i \theta^-\!(x) }e^{ i \theta^+\!(x)}e^{1/2\left[ \theta^-\!(x) , \theta^+\!(x) \right] }\right> = v \, e^{-1/2\left< \left[ \theta^+\!(x) , \theta^-\!(x) \right]\right> } \ ,\\
&= v \, e^{-1/2\left(\left<  \theta(0) \theta(0) \right>\right) } \ , \\
&= 0 \;\;\; \text{for d=2} \ ,
\end{aligned}
\end{equation}
where $\sigma(x)$ is the small massive norm perturbation and $v$ is the classical chosen vacuum. By using \eqref{2pointsCorrCFT2d} and translational symmetry, we notice that in two-dimensional spacetime, the VEV vanishes which makes inconsistent the initial hypothesis that $U(1)$ symmetry is spontaneously broken at quantum level. This concludes the sketch of the proof by contradiction.

It should be emphasised that it is the massless nature of the NG modes which leads to the logarithmic behaviour \eqref{2pointsCorrCFT2d}. If they were massive, we would not have a CFT and the two-point correlator would be an exponential decrease with an argument weighted by the mass of the considered particle. This explains why Coleman's theorem does not exclude the breaking of discrete symmetries or local symmetries in two dimensions. This is because such symmetries do not lead to Goldstone modes, i.e. to massless particles ``free'' in the IR. 

The non-relativistic scenario has been studied by Murayama and Watanabe \cite{Watanabe:2014fva} and is as well discussed in \cite{Beekman:2019pmi}. The idea remains the same except that they consider a generalisation of \eqref{nonRelLagCosetConstr} (at two dimensions there are additional symmetry-allowed terms). For the case where only type A NG modes are present, Coleman's theorem remains valid: no spontaneous symmetry breaking for $d=2$. This is consistent with the relativistic case since in relativistic theories all NG modes are type A. However, for the case with only type B NG modes, Coleman's theorem does not hold and it is is possible to have continuous global internal symmetry breaking. The mix case where both type A and type B NG modes interact is still an open question.   

Coleman's theorem can as well fail for specific relativistic theories. Theories with a large number $N$ of constituents are known to have ordered phases in the $N\to\infty$ limit \cite{Coleman:1974jh,Gross:1974jv}. It can be seen that the large quantum fluctuations are actually suppressed by a $1/N$ power \cite{Witten:1978qu}
\begin{equation}
\left< \phi(x)\phi(0) \right> \; \underset{\lvert x\rvert\rightarrow\infty}{\propto} \; \frac{1}{\lvert x\rvert^{1/N}}\underset{N\rightarrow\infty}{\longrightarrow} cst \ ,
\end{equation}
where $\phi$ is the fundamental field and its two-point correlator probes the ordered structure of the vacuum. This is precisely the case for theories which have a holographic dual. It was shown in \cite{Argurio:2016xih} that indeed $AdS_3$ holography allows for spontaneous symmetry breaking in its dual two-dimensional QFT.
 
Let us notice that large $N$ theories could a priori be seen as QFT curiosities rather than proper physical theories. So, the failure of Coleman's theorem is a rather axiomatic discussion. However, in the framework of holographic dualities, such models could describe sensible gravitational physics. 

\subsubsection{Mermin-Wagner-Hohenberg theorem}

A bit prior to Coleman work, a similar discussion has been done at finite temperature where the thermal fluctuations play a similar role as the quantum fluctuations on the parameter order. It is the Mermin-Wagner-Hohenberg theorem \cite{Mermin:1966fe, Hohenberg:1967zz}. It states that at finite temperature, no continuous spontaneous symmetry breaking can occur for $d\leq 3$ where $d$ is the spacetime dimension. Let us notice that the critical value for $d$ is more strict than the one for the zero temperature case (Coleman's theorem) which is consistent with the idea that now, both quantum fluctuations and thermal fluctuations add up in order to vanish the order parameter. 

In these lecture notes, we are not discussing thermal field theory, hence, we will not expand on how Goldstone physics fits with Mermin-Wagner-Hohenberg theorem. For the interested reader, a discussion on the possible NG modes in a thermal theory, at low spacetime dimension, is done (for example) in \cite{Watanabe:2014fva, Beekman:2019pmi}.

\subsection{Spontaneous breaking of spacetime symmetries}

Till now, we concentrate our study of the counting rules and of the classifications on the breaking of internal compact symmetry groups. We also discussed and argued that it is reasonable to think that we can extend the known results to the breaking of uniform symmetries. But concerning the breaking of spacetime symmetries, the analysis is much more involved. A generic counting rule for such symmetry breaking patterns is still unknown and represents a current active research topic. Because this field is not yet well established, we will remain sketchy in our discussion.  

\subsubsection{Spacetime symmetry specificities}

We can have a feeling why spacetime symmetry breaking is a complex problem. First of all, many of the spacetime symmetries are non compact -- e.g. dilatation, translations, boosts,... It means that the useful group theory properties we used so far are not anymore systematically verified. We did not expand much on it, but we have noticed that the coset construction can be related to differential geometry. The geometric study of coset spaces $G/H$ where $G$ is non compact is more involved. More broadly, taking $G$ totally general makes more difficult a generic classification of the NG modes, of all the possible terms in a symmetric-invariant EFT,... 

Furthermore, breaking spacetime symmetries usually means to work with a spacetime dependent background where before it was purely constant. Thus, the functional aspect of QFT is emphasised. Indeed, to find a stable vacuum, we minimise the energy. When we look for a constant solution, the energy can be seen as a function defined on a set of numbers (i.e. a real or a complex space). But, when we tolerate for spacetime modulated vacua, we are forced to consider the energy as it is, i.e. a functional. 

Another difficulty is that the effective Lagrangians are less constrained and thus, are more complicated. If we look at \eqref{relLagCosetConstr} and \eqref{nonRelLagCosetConstr}, we were able to write them in a compact form thanks to respectively Lorentz symmetry and rotation symmetry. The dispersion relations for the NG modes are therefore more involved with less constrained EFTs and can even be non analytic. We understand that the classification based on dispersion relations might have some flaws.   

Finally, Derrick theorem \cite{Rubakov:2002fi} suggest that we would need to have higher derivative terms in the fundamental theory to have stable and physical solitonic solutions. Of course, this should be qualified but it displays the tendency that even toy models are difficult theories. See for instance \cite{Musso:2018wbv, Musso:2019kii,Argurio:2021opr}. 

We just gave a gist of the technical difficulties associated to spacetime spontaneous symmetry breaking, but does it affect the counting rules we already know ? The answer is yes. In fact, even for relativistic fundamental theories, the number of NG modes can be reduced compared to the number of broken generators. Indeed, \cite{Salam:1969bwb,Volkov:1973vd} studied the spontaneous breaking of the conformal group down to the Poincaré group. They found that (for some cases) only one massless mode was present in the spectrum. It appears that it is the NG mode associated to the breaking of dilatation while the breaking of the special conformal transformations is not providing additional NG modes. A simpler example is the spontaneous breaking of translation and rotation symmetries to a discrete subset by a crystal lattice. With an explicit computation of the oscillation modes of the lattice, it can be noticed that the number of NG modes are linked to the breaking of translations and that the rotations do not provide additional massless excitations. We can intuitively understand this last result. 

\begin{figure}[h]
\centering{
\begingroup%
  \makeatletter%
  \providecommand\color[2][]{%
    \errmessage{(Inkscape) Color is used for the text in Inkscape, but the package 'color.sty' is not loaded}%
    \renewcommand\color[2][]{}%
  }%
  \providecommand\transparent[1]{%
    \errmessage{(Inkscape) Transparency is used (non-zero) for the text in Inkscape, but the package 'transparent.sty' is not loaded}%
    \renewcommand\transparent[1]{}%
  }%
  \providecommand\rotatebox[2]{#2}%
  \newcommand*\fsize{\dimexpr\f@size pt\relax}%
  \newcommand*\lineheight[1]{\fontsize{\fsize}{#1\fsize}\selectfont}%
  \ifx\svgwidth\undefined%
    \setlength{\unitlength}{269.79327746bp}%
    \ifx\svgscale\undefined%
      \relax%
    \else%
      \setlength{\unitlength}{\unitlength * \real{\svgscale}}%
    \fi%
  \else%
    \setlength{\unitlength}{\svgwidth}%
  \fi%
  \global\let\svgwidth\undefined%
  \global\let\svgscale\undefined%
  \makeatother%
  \begin{picture}(1,0.76595606)%
    \lineheight{1}%
    \setlength\tabcolsep{0pt}%
    \put(0,0){\includegraphics[width=\unitlength,page=1]{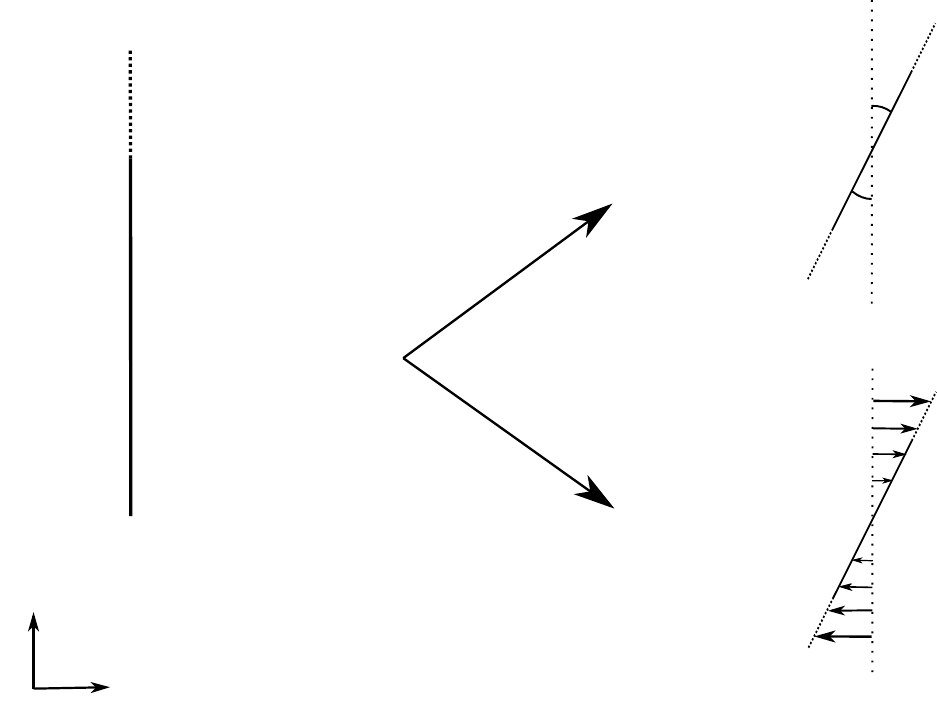}}%
    \put(0.94013868,0.65995787){\color[rgb]{0,0,0}\makebox(0,0)[lt]{\lineheight{1.25}\smash{\begin{tabular}[t]{l}$\theta$\end{tabular}}}}%
    \put(0.90428854,0.51269547){\color[rgb]{0,0,0}\makebox(0,0)[lt]{\lineheight{1.25}\smash{\begin{tabular}[t]{l}$\theta$\end{tabular}}}}%
    \put(0,0){\includegraphics[width=\unitlength,page=2]{trslSimRotFig.pdf}}%
    \put(0.36324753,0.49479554){\color[rgb]{0,0,0}\makebox(0,0)[lt]{\lineheight{1.25}\smash{\begin{tabular}[t]{l}$\sim \theta L_{xy}$\end{tabular}}}}%
    \put(0.38244678,0.25532138){\color[rgb]{0,0,0}\makebox(0,0)[lt]{\lineheight{1.25}\smash{\begin{tabular}[t]{l}$\sim y P_{x}$\end{tabular}}}}%
    \put(-0.00089852,0.11173408){\color[rgb]{0,0,0}\makebox(0,0)[lt]{\lineheight{1.25}\smash{\begin{tabular}[t]{l}$y$\end{tabular}}}}%
    \put(0.11138156,0.00339054){\color[rgb]{0,0,0}\makebox(0,0)[lt]{\lineheight{1.25}\smash{\begin{tabular}[t]{l}$x$\end{tabular}}}}%
    \put(0,0){\includegraphics[width=\unitlength,page=3]{trslSimRotFig.pdf}}%
  \end{picture}%
\endgroup%

\caption{In this cartoon, we illustrate how a modulated translation on a line can reproduce a global rotation of this line. This is the schematic reasoning providing that some NG modes associated to broken spacetime symmetries can be locked together.}
\label{trslSimRotFig}
}
\end{figure}

By looking at Figure \ref{trslSimRotFig}, we have an infinite straight rope disposed in a plane $Oxy$. By choosing this specific position/configuration, we spontaneously break translation symmetry in the $x-$direction and the rotation symmetry of the plane. We can observe that a global rotation acting on the rope is equivalent to a modulated action of the $x$-translation where the modulation is linear with $y$. If we extrapolate this information, we have that a local action of rotation on our rope can always be reproduced by a local action of $x$-translation. By definition, a NG mode is a spacetime modulation of the background in the direction of one of the spontaneously broken symmetries. Hence, the NG mode generated by the broken rotation is equivalent to the NG mode generated by the $x$-translation. So, unlike to the internal case, even before discussing dynamical conjugation between NG modes, we can already have locking between some broken directions ; and thus, a reduction of the number of independent NG modes. This instinctive reasoning has been formalised and talked through in \cite{Low:2001bw, Watanabe:2013iia, Brauner:2014aha}.

\subsubsection{Coset construction for spacetime symmetries}

The coset construction has been extrapolated to particular cases of spontaneous spacetime symmetries. In fact, we already mentioned that \cite{Salam:1969bwb,Volkov:1973vd} studied the spontaneous breaking of the conformal group down to the Poincaré group. Volkov in \cite{Volkov:1973vd} displayed a coset construction for an unspecified group $G$ spontaneously broken to a subgroup $H$ containing the Poincaré group. This particular extension to spacetime symmetries is explained in the lecture notes of Ogievetsky \cite{Ogievetsky:1974}. In this framework, it was noticed that, for some specific symmetry breaking patterns, it is possible to build invariant Lagrangians without requiring all the NG mode candidates. Hence, some of them might be non-physical or can be massive (e.g., the single dilaton associated to the breaking of dilatation symmetry and of special conformal transformation symmetries). The conditions when this situation occurs have been investigated by Ivanov and Ogievetsky in \cite{Ivanov:1975zq}. We provide below a schematic description of the prescription provided by \cite{Volkov:1973vd, Ogievetsky:1974, Ivanov:1975zq}, a more detailed review can be found in \cite{Penco:2020kvy}. The hypothesis of validity of this prescription are not yet well established. Thus, the following results should be taken with care and any application of the prescription to an extended case should carefully be checked. 

We consider a symmetry group $G$ which can include spacetime transformations. This group $G$ is spontaneously broken to a subgroup $H$ which contains the Poincaré group and other internal symmetries. We denote $X_a$ the broken generators, $P_\mu$ the unbroken translation generators and $T_A$ the remaining unbroken generators -- so, ${T_A}$ contains the Lorentz generators and some internal unbroken generators (notice the difference with Subsection \ref{Coset construction subsection}, where $T_A$ was denoting all the unbroken generators). The subgroup generated by $T_A$ is called $\tilde{H}$ and, on the contrary to the internal case, it is different from $H$. We ask for the additional relations
\begin{align}
& [X_a,T_A]= i f_{aA}^{\phantom{aA}b}X_b \ , \label{spacetimeRepbrokenunderH} \\
& [P_\mu,T_A]= i f_{\mu A}^{\phantom{\mu A}\nu}P_\nu \ .
\end{align}

The coset parametrisation is given by 
\begin{equation}
U\! \left(\pi(x),x\right)=e^{i x^{\mu} P_{\mu}}e^{i \pi^a(x) X_a} \ .
\label{spacetime coset parametrisation}
\end{equation} 
A possible intuition on why there is an additional $e^{i x^{\mu} P_{\mu}}$ factor compared to the internal case is that the coset parametrisation is built with the objects transforming non-linearly under the symmetry group. Now that the translations are explicitly listed in $G$, and because the coordinates $x^{\mu}$ transform as a shift (so, non-linearly) under the action of the translations, we intuitively understand that we have to introduce the $e^{i x^{\mu} P_{\mu}}$ factor in the coset parametrisation. The latter provides a supplementary term to the Maurer-Cartan 1-form
\begin{equation}
U(\pi)^{-1}\partial_{\mu}U(\pi) = -i \, \mathcal{A}_{\mu}^ A(\pi)\, T_A + i\, e_{\mu}^a(\pi)\, X_a + i \, r_\mu^\alpha(\pi) \, P_\alpha \ .
\label{spacetimeMauerCartan}
\end{equation}
The Maurer-Cartan 1-form preserves its covariant-like transformation properties despite the more involved transformation rules (the coordinates $x^\mu$ are, now, affected by the symmetries). We can thus, as usual, use $e_{\mu}^a(\pi)$ as our building block for an invariant Lagrangian. However, we have to pay attention that the partial derivative $\partial_\mu$ does not transform trivially anymore. The covariant derivative definition \eqref{covDerivCosetConstr} should include a new differential operator with appropriate transforming rules. This operator can be built thanks to $r_\mu^\alpha(\pi) \, P_\alpha $. We refer to \cite{Ogievetsky:1974,Penco:2020kvy} for its explicit construction. The same comment can be made on the measure $d^d x$ in the effective action. An invariant Lagrangian is not anymore sufficient to obtain an invariant theory since we have to ensure the invariance of the measure. An invariant volume form can be built from $r_\mu^\alpha(\pi) \, P_\alpha $.

\subsubsection{Inverse Higgs constraints}

Several examples in the literature (e.g. \cite{Salam:1969bwb,Volkov:1973vd}) emphasise that in particular cases with spacetime symmetry breaking, it is possible to get rid of some $\pi^a$ fields and still get an invariant theory. This happens when one (or several) of the Cartan form $e^a_\mu(\pi)$ depend linearly and additively on some of the $\pi^a$ fields, schematically it looks like
\begin{equation}
e^a_\mu(\pi)|_{(1)} \sim \partial_\mu \pi^a_{(1)} + \kappa \, \pi^a_{(2)} + \ldots \ ,
\label{IntuitionInvHiggs}
\end{equation}
where the numbers in brackets label different subsets of $\{\pi^a\}$ that will be expressed in terms of each other. The mathematical meaning of these subsets will be explained later. We understand that if we impose \eqref{IntuitionInvHiggs} to be zero, the obtained equation will be solvable and we can express the $\pi^a_{(2)}$ in terms of $\partial_\mu \pi^a_{(1)}$. Furthermore, this constraint will be consistent with the symmetries since $e^a_\mu(\pi)$ transforms covariantly and so, the constraint is symmetric invariant.

Such constraint bears the name of inverse Higgs constraint. The name comes from the case when we gauge the symmetry. In such a situation there are additional gauge fields and sometimes, we can impose a constraint like \eqref{IntuitionInvHiggs} to eliminate some of these gauge fields in terms of the NG candidates. Namely, it is sort of the Brout-Englert-Higgs mechanism in reverse. 

With additional hypotheses, Ivanov and Ogievetsky \cite{Ivanov:1975zq} established the conditions under which it is possible to impose an inverse Higgs constraint. The broken generators $X_a$ should be in a completely reducible representation of $\tilde{H}$ and we will denote with brackets the different multiplets $X_a^{(i)}$ (where the index $a$ now labelises the different generators inside the multiplet $(i)$)
\begin{align}
& [X_a^{(i)},T_A]= i f_{aA}^{\phantom{aA}b}X_b^{(i)} \ .
\end{align}
If
\begin{equation}
[P_\mu, X_a^{(i)}]\supset  f_{\mu a}^{\phantom{\mu a}b}X_b^{(j)} \ , \ f_{\mu a}^{\phantom{\mu a} b}\neq 0 \ ,
\end{equation}
we can impose the inverse Higgs constraint $e^a_\mu(\pi)|_{(j)}=0$ to eliminate the multiplet $\pi_a^{(i)}$ (the transformation rules under $\tilde{H}$ should also be consistent between $\pi_a^{(i)}$ and $\partial_\mu \pi_a^{(j)}$). 

After imposing the inverse Higgs constraint, the coset construction is equivalent to the one we would have done with $G'$, the reduced symmetry group where we subtracted the broken generators we got rid of. However, a trace of $G$ would remain in the relative numerical values of the coefficients of the Lagrangian. 

An example of inverse Higgs constraint could be the intuitive locking between some spacetime symmetries we illustrated at Figure \ref{trslSimRotFig}  \cite{Low:2001bw}. However, it is not always so obvious to know if an inverse Higgs constraint should be imposed. Indeed, it is not because we can do it that we should do it. It is an open question which is mainly dealt with by exploring several particular cases and by trying to deduce some objective criteria on which to decide to impose or not the inverse Higgs constraints. Here we provide some references which deal with this problematic and also, probe the coset construction for spacetime symmetries outside the hypothesis of these notes -- such as non-relativistic cases -- \cite{McArthur:2010zm, Nicolis:2013sga, Brauner:2014aha, Nicolis:2013lma, Endlich:2013spa }. To illustrate how this uncertainty on the inverse Higgs constraint impacts Goldstone physics, we emphasised at Subsection \ref{Goldstone physics at finite density} that there are additional massive NG modes at finite density which are not counted by Theorem \ref{NPBMW theorem}. A counting rule for such massive NG modes has been established in \cite{Nicolis:2013sga} by using the coset construction including a time translation symmetry breaking (coming from the chemical potential). Due to the current question mark we have on whether or not we have to impose some inverse Higgs constraints (so, whether or not we have a direct NG modes reduction), the obtained counting rule is an inequality rather than an equality.

\section{Take home message}

Goldstone Physics is a broad subject due to its main asset: it is a universal approach of the infra red physics which relies on symmetries. Low energy physics is mainly our daily life surroundings, which makes it observable by definition. Therefore, Goldstone Physics is a formal description of physics but it can almost straightforwardly provide material for phenomenology and for experiments. This explains the large scope of this area of science.   

Goldstone physics is two folds. First, it is the study of the infra red spectrum from the perspective of symmetry-originated modes. When spontaneous symmetry breaking occurs, given additional not too restrictive conditions, the spectrum will contain massless particles (NG modes) and light particles (pseudo NG modes). Theorems exist to provide us information on their number and their characteristics. An active research direction is to generalise and to enrich these theorems.

Second, Goldstone physics has the aim to build the most general shape of an effective field theory for a given symmetry pattern. To do so, it uses several building methods, consequently, it helps to improve these methods. 

In conclusion, knowing both the spectrum content and the shape of the theory at low energy entirely fix the dynamics of the infra red physics. The approach is mainly based on symmetry concepts which are model independent. Hence, the outcomes are generic, which leads to universality. 

\section*{Acknowledgments}

The author would like to thank the organisers of the XVII Modave summer school in mathematical physics for their invitation to give this lecture and for creating such an enjoyable atmosphere during the school. This acknowledgment goes as well to the participants for their enthusiasm and for their interesting questions and comments. 

A special thank you goes to Riccardo Argurio and to Daniele Musso for their careful proofreading and comments on the first draft of these notes. It has led to a significant improvement of the manuscript.

\bibliographystyle{utphys}
\bibliography{GoldstonePhysEFTBiblio} 

\end{document}